\documentclass[traditabstract]{aa} 
\usepackage{graphicx}
\usepackage{txfonts}
\usepackage{mathtools}
\usepackage{numprint}
\usepackage{color}
\usepackage{placeins}
\npdecimalsign{.}
\nprounddigits{0}
\newcolumntype{M}[2]{>{\hfill}n{#1}{#2}<{\hfill}}
\newcolumntype{N}[2]{>{\hfill}n{#1}{#2}<{\hfill}}
\newcommand{\ratioo} {N({\rm H}_2) / I_{\rm CO}}
\newcommand{\Htwo} {\rm H_{2}}
 
\usepackage[colorlinks,linkcolor=blue,anchorcolor=blue,citecolor=blue]{hyperref}

\def\la{\lower.5ex\hbox{$\; \buildrel < \over \sim \;$}}
\def\ga{\lower.5ex\hbox{$\; \buildrel > \over \sim \;$}} 
\begin{document}
   \title{Predicting CO and dust emission of star-forming galaxies}

   \subtitle{Extension to star-forming low-mass and dwarf galaxies}

   \author{B.~Vollmer\inst{1}, J.~Braine\inst{2}, M.~Soida\inst{3} and P.~Gratier\inst{2}}

   \institute{Universit\'e de Strasbourg, CNRS, Observatoire Astronomique de Strasbourg, UMR 7550, 67000 Strasbourg, France \and
     Laboratoire d'Astrophysique de Bordeaux, Univ. Bordeaux, CNRS, B18N, all\'e Geoffroy Saint-Hilaire, 33615 Pessac, France \and
     Astronomical Observatory, Jagiellonian University, ul. Orla 171, 30-244 Krak\'ow, Poland
     }

   \date{Received ; accepted }


  \abstract
      {How do Dwarf Galaxies differ from spirals? Does star formation produce radio and far-infrared emission in the same way as in spiral galaxies?  Radio, FarIR, and CO emission
      depend on gas density, temperature, magnetic field strength, and metallicity. The radio-FarIR correlation and Schmidt-Kennicutt relation characterize the links for Milky Way-like galaxies but do they hold for smaller objects, with different morphologies?
        Here we extend our previous work on the IR, line, and radio emission 
        of local and high-z galaxies to local star-forming low-mass and dwarf galaxies. The calculation of the cosmic ray (CR)  
        densities were improved compared to the previous version of the model.
        The CR ionization rate we found for the different galaxy samples is higher by a factor of three than
        for the solar neighborhood. This means that the mean yield of low-energy CR particles 
        three times higher in external galaxies than was observed by Voyager~I.
        The dependence of the $\ratioo$ factor on the metallicity and stellar mass are calculated by the model.
        The weaker CO emission from low-metallicity galaxies is due to the large amount of (CO-dark) $\Htwo$ surrounding the regions where CO is not photo-dissociated.
      Within our model framework, star-forming low-mass and dwarf galaxies follow the radio--IR correlation.
      }


   \authorrunning{Vollmer et al.}

   \maketitle
   \nolinenumbers
%

\section{Introduction\label{sec:introduction}}

The radio--infrared correlation and the Schmidt-Kennicutt relation between the molecular gas mass and the star formation rate are well established
in star-forming galaxies with stellar masses higher than $\sim 10^9$~M$_\odot$ (see, e.g., Yun et al. 2001, Bell 2003, Molnar et al. 2021, Kennicutt \& De Los Reyes 2021,
Schinnerer \& Leroy 2024). Whether these relations also hold in star-forming low-mass and dwarf galaxies is still matter of debate. The physical conditions in these
low-mass and low-metallicity galaxies are distinct from those of their high-mass counterparts. Observationally, the quality of using
the radio continuum, IR emission, and CO emission as tracers of the different ISM components depends on many factors, such as gas density, temperature,
magnetic field strength, and metallicity. 

Detecting CO in dwarf galaxies has been notoriously difficult (e.g., Leroy et al. 2009a, Cormier et al. 2014) because of the faintness of the CO emission.
Low metallicity environments imply lower C and O abundances and low dust-to-gas ratios (Draine et al. 2007). This decreases the H$_2$ abundance
since H$_2$ molecules form on dust surfaces. Dust provides much of the far-UV shielding that is necessary to prevent CO, which generally provides little self-shielding,
from photodissociating (Bolatto et al. 2013). Model predictions point to a CO deﬁciency in low-metallicity star-forming galaxies as being
due to a decrease in dust-shielding, which leads to strong photo-dissociation of CO by the intense UV radiation ﬁelds that are generated in
the star-forming regions (e.g., Wolﬁre et al. 2010). H$_2$, on the other hand, photodissociates via absorption of FUV photons, which, for moderate A$_{\rm V}$,
can become optically thick, allowing H$_2$ to become self-shielded from photodissociation (e.g. Gnedin \& Draine 2014), leaving a potentially signiﬁcant
reservoir of H$_2$ existing outside of the CO-emitting core, which is commonly called CO-dark molecular gas.
As a consequence, the CO-H$_2$ conversion factor strongly increases with decreasing metallicity (e.g., Ramambason et al. 2024).
Within the CO-dark photodissociation regions [C{\sc ii}]$158\mu$m emission can be used as a tracer of molecular hydrogen (e.g., Madden et al. 2020).

Many authors have reported that the star formation law derived for spiral galaxies does not hold in dwarf galaxies, suggesting the existence of different
regimes of star formation for different galaxy types. The SF surface densities of dwarf galaxies tend to be lower than the values expected from the
Schmidt-Kennicutt relation (Wyder et al. 2009; Gatto et al. 2013; Roychowdhury et al. 2015, 2017; de los Reyes \& Kennicutt 2019 but see Teich et al.
2016 for a different conclusion). These studies are based on the atomic gas surface density.

Using  CO emission to trace H$_2$ is difficult in low-metallicity, generally low-mass, galaxies because the CO emission 
per H$_2$ mass is much lower, meaning that the CO-H$_2$ conversion factor $\alpha_{\rm CO}$ is higher than in large spirals.
Different methods to estimate $\alpha_{\rm CO}$ in dwarf galaxies yielded
discordant values (e.g. Leroy et al. 2011; Bolatto et al. 2013; Hunt et al. 2015; Amor\'{i}n et al. 2016; Madden \& Cormier 2019), hence it is difficult to estimate the $\Htwo$ content of low-metallicity galaxies.  Cloud-resolved studies of the small local group galaxies M\,33 and NGC\,6822 suggest that molecular gas is converted in stars more quickly (i.e. a lower $\Htwo$ depletion time (Gardan et al. 2007, Gratier et al. 2010, Gratier et al. 2017) than in large spirals.
This is to be expected because of the weaker stellar winds (Dib 2011) and higher density for conversion of H{\sc i} to $\Htwo$ at low metallicities.

Radio continuum emission is emitted by ionized gas, as well as through cosmic rays (CR) interacting with magnetic ﬁelds (synchrotron emission).
The radio continuum emission at centimeter wavelengths is generally produced
by a combination of free–free emission from thermal electrons, and non-thermal synchrotron emission from relativistic electrons
spiraling around magnetic ﬁelds (e.g. Condon 1992). Increased ionizing UV photon production, low
metallicity,  and reduced dust absorption can increase the ionization of the gas, and, hence, the thermal emission component.
This effect is counteracted by increased ionizing photon escape from the galaxy (e.g. Fernandez \& Shull 2011; Benson, Venkatesan \& Shull 2013;
Leitherer et al. 2016). Together with the SFR, the gravitational potential determines the CR loss versus injection budget: the star
formation surface density threshold to drive winds and outﬂows, which can be responsible for CR advection and escape from the galaxy, is lower
in smaller gravitational potential wells. It is thus not obvious that the radio-infrared correlation holds for star-forming low-mass and dwarf galaxies.

In this article we apply the model of IR, molecular line, and radio continuum emission presented in Vollmer et al. (2017, 2022, and 2025) to
the sample of star-forming low-mass and dwarf galaxies presented in Leroy et al. (2008 and 2009b).
We investigated if the model can reproduce the available integrated IR, CO, and radio continuum observations. Since the CO-H$_2$ conversion factor is
calculated by the model, the mechanism for its increase with decreasing metallicity can be identified within the model framework.

The article is structured in the following way: the new galaxy sample and the model are introduced in Sect.~\ref{sec:observations} and Sect~\ref{sec:model}.
The model results are presented and compared to available observation in Sect.~\ref{sec:modelresults}. 
The implications on the CO conversion factor, the CO-dark gas, the thermal fraction of the radio continuum emission, the star formation law,
and the viscous gas transport are discussed in Sect.~\ref{sec:discussion}, followed by our conclusions (Sect.~\ref{sec:conclusions}).

\section{The star-forming low-mass and dwarf galaxy sample\label{sec:observations}}

We apply our models to the the galaxy sample presented in Leroy at al. (2008). In their study
these authors used H{\sc i} maps from the H{\sc i} Nearby Galaxy Survey (THINGS) and CO maps measured by the HERA CO-Line Extragalactic Survey
and the Berkeley-Illinois-Maryland Association Survey of Nearby Galaxies. Their galaxy sample was divided in two parts, high- and low-mass galaxies.
Models of galaxies with stellar masses higher than
$10^{10}$~M$_{\odot}$ were presented in Vollmer et al. (2017, 2022, and 2025). The galaxies of masses $10^7 \le M_* \le 10^{10}$~M$_{\odot}$
are presented here (Table~\ref{tab:sample}). The rotation velocities range between $50$ and $130$~km\,s$^{-1}$.
The metallicities vary between 12+log(O/H)$=7.5$ to $8.5$ (Fig.~\ref{fig:metallicities}). The CO data were provided by the HERACLES survey (Leroy et al. 2009).
\begin{table*}
      \caption{Galaxy sample}
         \label{tab:sample}
      \[
       \begin{tabular}{lccccccccccc}
         \hline
         name & dist & type & $r_{25}$ & $v_{\rm flat}$ & $l_{\rm flat}$ & log($M_*$) & log($M_{\rm HI}$) & SFR & $l_*$ & $\rho_{0,{\rm DM}}$ & $R_{0,{\rm DM}}$ \\
         & (Mpc) & & (kpc) & (km\,s$^{-1}$) & (kpc) & (M$_{\odot}$) &   (M$_{\odot}$) & (M$_{\odot}$yr$^{-1}$) & (kpc) & ($10^5$~M$_{\odot}$pc$^{-3}$) & (kpc) \\
         \hline
         DDO~154 & 4.3 & Irr & 1.2 & 50 & 2.0 & 7.1 & 8.7 & 0.005 & 0.8 & 2.3 & 1.7 \\
        Ho~{\sc i} & 3.8 & Irr & 1.8 & 53 & 0.4 & 7.4 & 8.3 & 0.009 & 0.8 & 1.3 & 0.15 \\
        Ho~{\sc ii} & 3.4 & Irr & 3.7 & 36 & 0.6 & 8.3 & 8.9 & 0.048 & 1.2 & 0.5 & 0.2 \\
        IC~2574 & 4.0 & Irr & 7.5 & 134 & 12.9 & 8.7 & 9.3 & 0.070 & 2.1 & 15.0 & 10.0 \\
        NGC~4214 & 2.9 & Irr & 2.9 & 57 & 0.9 & 8.8 & 8.7 & 0.107 & 0.7 & 4.0  & 2.0 \\
        NGC~2976 & 3.6 & Sc & 3.8 & 92 & 1.2 & 9.1 & 8.3 & 0.087 & 0.9 & 10.0 & 2.0 \\
        NGC~4449 & 4.2 & Irr & 2.8 & 97 & 1.0 & 9.3 & 9.2 & 0.371 & 0.9 & 6.0 & 1.0 \\
        NGC~3077 & 3.8 & Sd & 3.0 & 100 & 1.0 & 9.3 & 9.1 & 0.086 & 0.7 & 11.0 & 2.0 \\
        NGC~7793 & 3.9 & Scd & 6.0 & 115 & 1.5 & 9.5 & 9.1 & 0.235 & 1.3 & 9.0 & 1.5 \\
        NGC~2403 & 3.2 & SBc & 7.3 & 134 & 1.7 & 9.7 & 9.5 & 0.382 & 1.6 & 11.0 & 1.5 \\
        NGC~925 & 9.2 & SBcd & 14.2 & 136 & 6.5 & 9.9 & 9.8 & 0.561 & 4.1 & 13.0 & 5.0 \\
        \hline
        \end{tabular}
      \]
\end{table*}

In addition, we applied our model to the galaxy samples described in Vollmer et al. (2017, 2025):
  local luminous infrared galaxies (LIRGs; Fisher et al. 2019), local starbursts (ultraluminous infrared galaxies; Downes \& Solomon 1998),
  intermediate-redshift ($z \sim 0.3$-$0.5$) main sequence
  galaxies (PHIBSS2; Freundlich et al. 2019), high-z main sequence (PHIBSS, Tacconi et al. 2013) and starburst galaxies (Genzel et al. 2010).

\section{The model \label{sec:model}}

Galactic gas disks are modeled as turbulent clumpy star-forming accretion disks. The details of the gas disk model are described in Appendix~\ref{sec:dgasdisk}.
The turbulent nature of the ISM is taken into account by
applying scaling relations for the gas density and velocity dispersion to gas clouds of given sizes. The scaling relations change when the clouds become self-gravitating.
The cloud temperature is calculated via the balance of heating and cooling. In this way, the gas density, temperature, and
velocity dispersion were determined for each cloud. The molecular abundances depend on the density, temperature,
and CR ionization rate. The CR ionization rate was calculated by using a steady-state model (Pohl 1993) calibrated by the CR ionization in the
solar neighborhood. The molecular abundances were derived from calculations by the time-dependent gas-grain code Nautilus (Hersant et al. 2009, Ruaud et al. 2016).
The molecular abundances, gas density, velocity dispersion, and temperature served as input for the calculation of the molecular line emission.
The radio continuum emission was calculated via a steady-state model calibrated by the observed IR-radio correlations.
We performed simulations to predict the emission at many different wavelengths of observed galaxies.

The model simultaneously calculates radial profiles of the gas surface density $\Sigma$, 
the SF surface density $\dot{\Sigma}_*$, the gas velocity dispersion, and the volume-filling factor, all of which are large-scale properties,
and at small scales $f_{\rm mol}$ and the IR SED and molecular line emission. 
The molecular line emission was calculated via RADEX (van der Tak 2007).
For the details of our analytical model, we refer to Appendix~A of Liz\'ee et al. (2022).
The stellar mass profile is given by the S\'ersic fit to the observations, so that the calculations are of the gas distribution and kinematics.  

The rotation curve and the surface density profile of the stellar disk served as model inputs.
For the low-mass and dwarf galaxies we added a dark matter halo with a cored isothermal density profile
\begin{equation}
\rho(R)=\frac{\rho_0}{(R+R_0)^2}\ ,
\end{equation}
where we adapted $\rho_0$ and $R_0$ to obtain, together with the stellar disk, the observed rotation curve given by Leroy et al. (2008).
The central densities and core radii are given in Table~\ref{tab:sample} for the different galaxies.
The halo is also taken into account for the vertical hydrostatic equilibrium.

The model contains three main free parameters:
(i) the Toomre parameter $Q=\frac{c_s \kappa}{\pi G \Sigma}$ of the gas, where a $Q<1$ disk is gravitationally unstable
(ii) the external accretion rate
$\dot{M}$, where an increasing $\dot{M}$ at constant $Q$ increases $v_{\rm turb}$, and (iii) $\delta = l_{\rm driv}/l_{\rm cl}$, where $l_{\rm cl}$ is the cloud size
scale and $l_{\rm driv}$ is the driving length of the turbulence.
The mass accretion rate and $\delta$ are assumed to be constant. Given the Toomre parameter, mass accretion rate, rotation curve,
and stellar mass surface density and velocity dispersion, the large-scale accretion disk model yields the gas density, disk height, gas surface density,
gas turbulent velocity dispersion, and star formation surface density. The turbulent nature of the interstellar medium (ISM) is taken into account by
applying scaling relations for the gas density and velocity dispersion to gas clouds of given sizes. The scaling relations change when the clouds become self-gravitating.
The cloud temperature is calculated via the balance of heating and cooling. In this way, the gas density, temperature, and
velocity dispersion were determined for each cloud. The molecular abundances depend on the density, temperature,
and CR ionization rate. The molecular abundances, gas density, velocity dispersion, and temperature served as input for the calculation of the molecular line emission.
The radio continuum emission was calculated via the steady-state model of Pohl (1993) calibrated by the observed IR-radio correlations.
For each radius $R$, the
model yields the SF surface density $\dot{\Sigma}_{*}$, the H{\sc i} emission, and the IR continuum and molecular line emission for a given set of
$Q$, $\dot{M}$, and $\delta$. Liz\'ee et al. (2022) showed that these radial profiles do not significantly depend on $\delta$
for $2 \le \delta \le 10$. We thus set $\delta=5$. 
The constant that links the rate of energy injection by supernovae to the SFR was set at $\xi = 9.2 \times 10^{-8}$~(pc\,yr$^{-1}$)$^2$.
The details of the gas disk model are described in Appendix~A of Vollmer et al. (2025).

Following Vollmer et al. (2017), our leaky-box model used an effective yield, meaning that enriched gas can escape and be accreted from the circumgalactic medium.
This led to a gas metallicity of
\begin{equation} \label{eq:zzodot}
Z/Z_{\odot}=(0.3~{\rm yr\,M_{\odot}pc^{-3}})/\alpha
\end{equation}
with
\begin{equation}
\label{eq:metulirgs}
\alpha=\left( \ln\left(\frac{\Sigma_{*}+\Sigma}{\Sigma}\right)\right)^{-1}\ ,
\end{equation}
where $\Sigma_*$ is the stellar mass surface density.
The normalization of Eq.~\ref{eq:zzodot} is half that of the high-mass galaxies (Vollmer et al. 2017, 2022, and 2025)
to reproduce the integrated metallicities given by McQuade et al. (1995), Kennicutt et al. (2011), and Madden et al. (2013).
The observed metallicities of all galaxies more massive than Ho~{\sc i} are reproduced within $1\sigma \la 0.1$~dex
(Fig.~\ref{fig:metallicities}). Those of DDO~154 and Ho~{\sc i} are underestimated by $0.3$~dex and $0.5$~dex, respectively.

To determine the chemical abundances, we used seven grids calculated by the time-dependent gas-grain code Nautilus (Hersant et al. 2009, Ruaud et al. 2016)
for CR ionization rates of
$(0.1,\,0.3,\,1.3,\,3.0,\,6.5,\,30,\,130,\,650,\,6500) \times 10^{-17}$~s$^{-1}$ and varied the time, gas density, and gas temperature.
The chemical abundances were calculated by interpolating the four-dimensional Nautilus grid for a given CR ionization rate (Sect~\ref{sec:crion})
and solar metallicity.

The abundances scale with the model metallicity. Since Nautilus is designed to calculate the molecular abundances in gas with 
$n \ga 10^3$~cm$^{-3}$, the Nautilus abundances were extrapolated to lower densities. This extrapolation only affects the CO emission
of the local non-starburst galaxies, particularly the low-mass and dwarf galaxies. We validated our approach by recalculating the CO(1-0)
emission of these galaxies using the simple phenomenological model of CO formation as the immediate descendant of quiescently-recombining
HCO$^+$ (Eq.~4c of Liszt 2007\footnote{Luo et al. (2023) combined absorption observations of HCO$^+$, and absorption and emission observations of CO to
calculate their column densities in the diffuse ISM.
The HCO$^+$/CO column density ratio is about constant within a factor of two to three for $n \ga 30$~cm$^{-3}$ (Table~2 of Luo et al. 2023).}).
Changes in the model CO(1-0) fluxes are less than $\sim 10$\,\% for all galaxies, except DDO~154 where the recalculated
flux is $\sim 40$\,\% lower than the initial flux. These changes are well within the uncertainties of the model.
In addition, photo-dissociation of H$_2$ and CO molecules (CO-dark gas) in the outer envelopes of gas clouds following the approach of
Krumholz et al. (2008, 2009) and Wolfire et al. (2010) is included in our model (see Appendix~\ref{sec:dissociation}).

The line emission was calculated with RADEX using the inputs as calculated above.
The total emission is highly dependent on the gas mass surface densities (and hence filling factors). These come from the local gas densities, and the prescription
which decides whether gas is atomic or molecular based on the competition between the molecular formation time and the freefall time. The details are described in
Sect.~2.1.1 and 2.4 of Vollmer et al. (2017). 

\subsection{Cosmic-ray ionization rate \label{sec:crion}}

Cosmic rays with energies above $1$~MeV interact with atoms and molecules of the interstellar medium.
These high-energy particles can directly ionize a species. This direct process is dominant for H$_2$ , H, O, N, He,
and CO. Electrons produced in the direct process can cause secondary ionization before they are thermalized.
In addition to secondary ionization, the electrons produced in direct CR ionization can electronically excite
molecular and atomic hydrogen. The radiative relaxation of H$_2$ (and H) produces UV photons that ionize and
dissociate molecules. Whereas direct ionization is dominant for CO, secondary ionization and dissociation are dominant for HCN.
The CR ionization rate thus affects the chemical abundances.

As in Vollmer et al. (2025) we calculated the CR ionization rate, which was then injected into 
Nautilus, which uses the total H$_2$ CR ionization rate $\zeta_{{\rm H}_2}$ as input (Wakelam et al. 2012).
The CR nucleons (mainly protons) lose energy by the combined effects of ionization and Coulomb losses, adiabatic
deceleration, and pion production. To calculate the local (solar neighborhood) CR density spectra of electrons and protons,
we followed Pohl (1993). The normalization factors of the proton spectrum was chosen such that the observed local CR intensity,
averaged over a solid angle, $J(E)$, was recovered (Fig.~3 of Vollmer et al. 2025).
To calculate the CR ionization rate at a given distance $R$ from the center of a galaxy, we set $n_{\rm H}$ to the
gas density of the analytical galaxy model (Appendix~\ref{sec:dgasdisk}), use the interstellar radiation field $U$ and the magnetic field strength
of the galaxy model at $R$, and multiply the proton and electron spectra by $\dot{\rho}_*$ of the galaxy model.
In the Vollmer et al. (2025) the normalization was done with the SF surface density $\dot{\Sigma}_*$.

We had a closer look at the CR ionization rates at the scale lengths of the stellar disks.
On average, the new normalisation led to CR ionisation rates about five times higher in local main-sequence, luminous infrared and starburst galaxies.
The increase of the CR ionization rates was about a factor of two to three in the intermediate-redshift LIRGs and high-redshift main-sequence galaxies, whereas it
was negligible in the high-z starburst galaxies. 
However, this only led to a maximum increase in CO(1-0) line emission by $\la 50$\,\% across all sample galaxies.
The increase of the CO line emission is greater for higher transitions. The CO(4-3) line emission increased by about a factor of $1.5$ to $2$.
On the one hand, an increased CR ionization rate leads to a lower CO abundance. On the other hand, it increases the CR gas heating.
As CO emission increased with increasing CR ionization rate, the effect of the gas heating dominates.
  
\subsection{Radio continuum \label{sec:radcont}}

CR electrons are accelerated to relativistic velocities in expanding supernova (SN) shells.
The exponent of the CRe injection spectrum is $q=2.3$ as expected for superbubbles created
by multiple SN remnants (Vieu et al. 2022). For relativistic electrons, ionization and Coulomb losses, non-thermal bremsstrahlung, adiabatic deceleration,
inverse Compton losses, and synchrotron emission have to be taken into account (e.g., Lacki et al. 2010).
The normalization factor of the electron spectrum was chosen such that the observed local CR intensity,
averaged over a solid angle was recovered (Fig.~3 of Vollmer et al. 2025).
To calculate the CR ionization rate at a given distance $R$ from the center of a galaxy, we set the gas density $n_{\rm H}$ to the
gas density of the analytical galaxy model, use the interstellar radiation field $U$ and the magnetic field strength
of the galaxy model at $R$, and normalize the equilibrium spectrum of CR protons and electrons with the SF volume density $\dot{\rho}_*$ of the
galaxy model.

Following Vollmer et al. (2022), we assumed a stationary CR electron density distribution ($\partial n/\partial t$=0). 
The CR electrons are transported into the halo through diffusion or advection
where they lose their energy via adiabatic losses or where the energy loss through synchrotron emission is so small that the
emitted radio continuum emission cannot be detected. Furthermore, we assumed that the source term of CR electrons is
proportional to the SFR volume density $\dot{\rho}_*$. For the energy distribution of the CR electrons, the standard assumption is a power law
with index $q$, which leads to a power law of the radio continuum spectrum with index $-(q-1)/2$ (e.g., Beck 2015).
The details of the radio continuum model are described in Appendix~B of Vollmer et al. (2025).

As a supplementary step we replaced Eq.~B.11 of Vollmer et al. (2025) by calculating the radio continuum emission according to chapter~5 of
Condon \& Ransom (2016) and chapter~3 of Pacholczyk (1970). For the CR energy loss terms we used the formalism of Pohl (1993) with the
energy loss time scales of Vollmer et al. (2022).
The new method has the advantage that the normalization 
of the CR ionization rate and yield of low-energy CR particles are explicitly set
within the model framework; they are  three times higher than was observed by Voyager~I and very close to the results presented in Vollmer et al. (2025).

\section{Model result \label{sec:modelresults}}

The model calculates the IR spectral energy distributions (SED), CO emission, and the radio continuum SEDs.
These are compared to available observations in Sects.~\ref{sec:IR}, \ref{sec:CO}, and \ref{sec:RC}, respectively.

\subsection{Infrared emission \label{sec:IR}}

Following Vollmer et al. (2017), we extracted all available photometric data points for our galaxy samples from the CDS VizieR
database\footnote{\tt http://vizier.u-strasbg.fr/viz-bin/VizieR} for a direct comparison between the model and the observed dust SED. Since the flux densities
were determined within different apertures, we only took the highest flux densities for
a given wavelength range around a central wavelength $\lambda_0$ ($0.75 \leq \lambda/\lambda_0 \leq 1.25$).
In this way, only the outer envelope of the flux density distribution was selected. We also separately show the Herschel measurements from Dale et al. (2012).
Since our dust model does not include stochastically heated small grains and PAHs, the observed IR flux densities
for $\lambda \la 50~\mu$m cannot be reproduced by the model. 

\begin{figure}
  \centering
  \resizebox{\hsize}{!}{\includegraphics{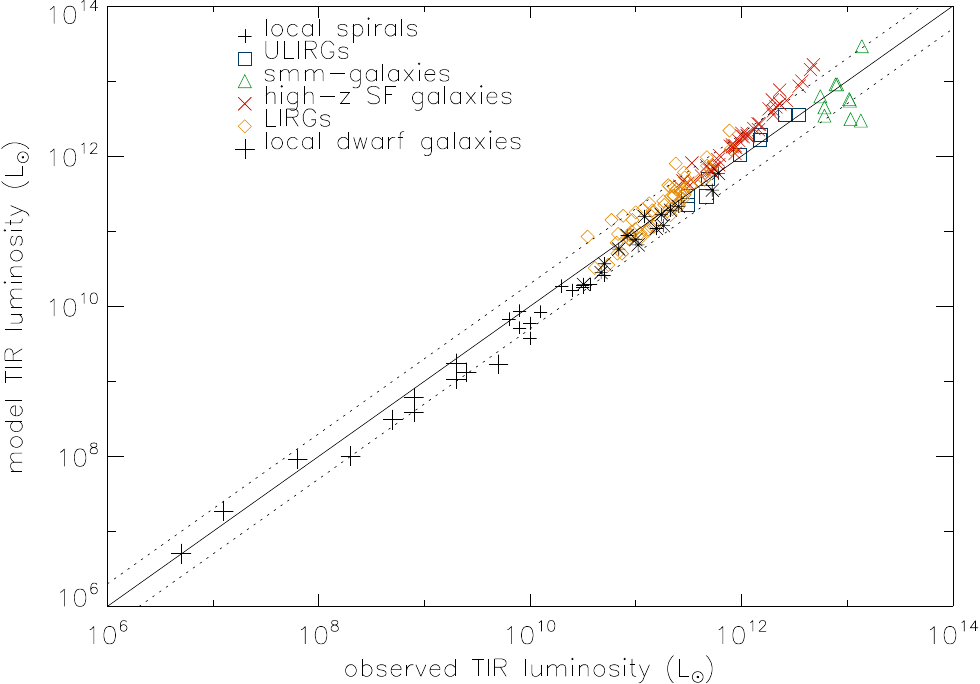}}
  \caption{Model TIR luminosity as a function of the observed TIR luminosity. The dotted lines show a factor two above and below the equality. 
  \label{fig:plots_HCNCO_dwarfs_1}}
\end{figure}

We assumed a dust mass absorption coefficient of the following form:
\begin{equation}
\label{eq:kappa}
\kappa(\lambda)=\kappa_0\,(\lambda_0/\lambda)^{\beta}\ ,
\end{equation}
with $\lambda_0=250~\mu$m, $\kappa_0=0.48~{\rm m}^2{\rm kg}^{-1}$ (Dale et al. 2012), and a gas-to-dust ratio of $GDR=M_{\rm gas}/M_{\rm dust}=\frac{Z_{\odot}}{Z} \times 100$ 
(including helium; R{\'e}my-Ruyer et al. 2014). The exponent was $\beta=1.5$ for the star-forming low-mass and dwarf galaxies.
The corresponding model and observed $10~\mu$m to $1000~\mu$m luminosities (hereafter TIR) and dust IR SEDs are presented in Figs.~\ref{fig:plots_HCNCO_dwarfs_1} and
\ref{fig:IRspectra_dwarfsZ}. The observed TIR luminosities are taken from Kennicutt et al. (2011), Dale et al. (2012), Rujopakarn et al. (2013), Cormier et al. (2015),
R\'emy-Ruyer et al. (2015), and Aniano et al. (2020). Overall, the model and observed  TIR luminosities agree within a factor of two.
Most of the model TIR luminosities of the local main sequence galaxies (small and large crosses in Fig.~\ref{fig:plots_HCNCO_dwarfs_1}) are underestimated by about $50$\,\%.

As for the TIR luminosities, the observed IR SEDs are reproduced by the model within a factor of two. The model IR SEDs of Ho~{\sc i} and Ho~{\sc ii} are
somewhat overestimated, whereas that of NGC~925 is somewhat underestimated. Since the uncertainty of our model results is about a factor of two (Vollmer et al. 2017),
we did not try to adjust the SFR of the sample galaxies to obtain model results closer to observations.
We fitted modified Planck functions with $\beta=1.5$ to the model dust IR SEDs to derive the dust temperatures.
The modified Planck functions are shown as dashed red lines in Fig.~\ref{fig:IRspectra_dwarfsZ}.
The resulting mean temperatures of the large grain population lie within $13$ and $23$~K with a mean of $19 \pm 3$~K.

\subsection{CO emission \label{sec:CO}}

\begin{figure}
  \centering
  \resizebox{\hsize}{!}{\includegraphics{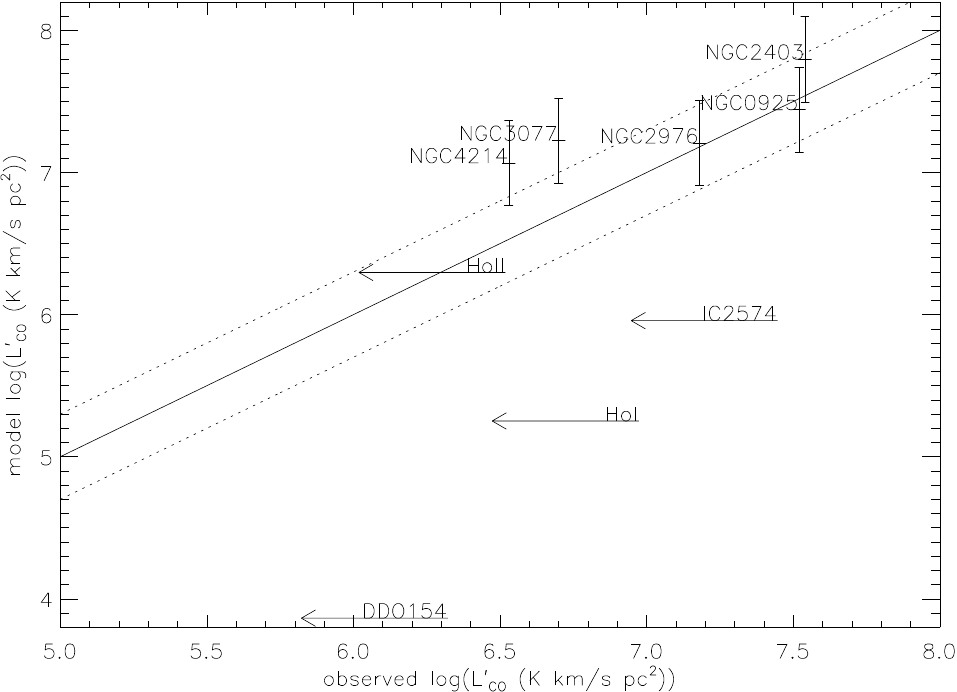}}
  \caption{Model CO luminosity as a function of the observed CO luminosity for the star-forming low-mass and dwarf galaxies. The dotted lines show a factor two above and below the equality. 
  \label{fig:COcomparison}}
\end{figure}

Since the CO emission of star-forming low-mass and dwarf galaxies is notoriously low, as stated in Sect.~\ref{sec:introduction}, the CO emission has only
been detected in five out of 11 galaxies (Leroy et al. 2009, 2022). For four galaxies upper limits are available.
The model and observed CO(2-1) luminosities and upper limits are shown in Fig.~\ref{fig:COcomparison}.
The model overpredicts the observed CO luminosities by about $50$\,\%. The model CO luminosities are consistent with the observed upper limits.
Ho~{\sc ii} probably is close to the detection limit. We also calculated models with half the gas metallicities. These models resulted in about
$50$\,\% lower CO luminosities and reproduce the observed CO luminosities within a factor of two, which is within the model uncertainties
(Vollmer et al. 2017).

\subsection{Radio continuum emission \label{sec:RC}}

The model radio continuum emission from $150$~MHz to $1.4$~GHz was calculated using the framework presented in Sect.~\ref{sec:radcont}.
The radio SEDs of the star-forming low-mass and dwarf galaxies are presented in Fig.~\ref{fig:radiospectra_dwarfs}. As for the IR SEDs, we also show measurements
available in the CDS/VizieR database. Overall, the model reproduced the observed radio SEDs within a factor of two. Only for NGC~4214, the model overestimated the
radio flux densities by about a factor of three. In agreement with the available observations, the slope of the radio SEDs significantly flattens at high
frequencies. This is caused by an increase of the fraction of thermal emission towards higher frequencies (see Sect.~\ref{sec:thermalf}).

\begin{figure*}
  \centering
\resizebox{\hsize}{!}{\includegraphics{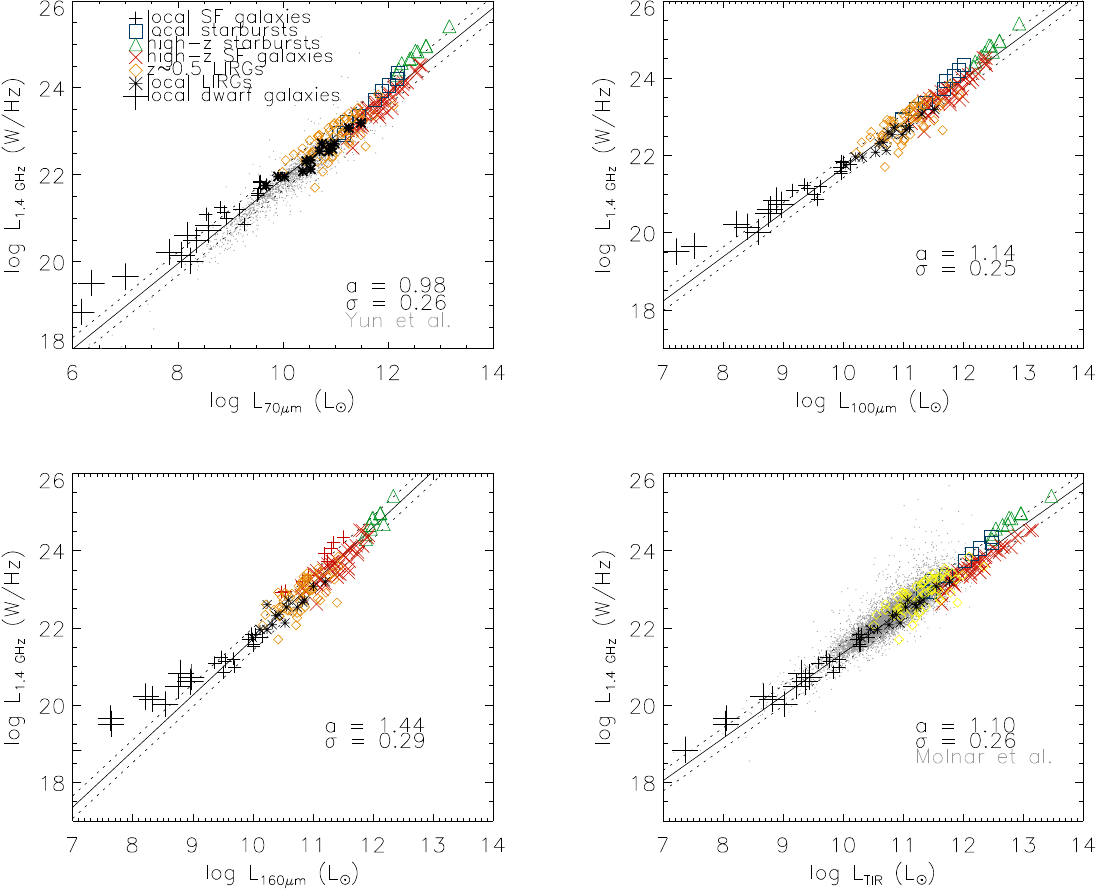}}
\caption{IR-radio correlations. Upper left: $70$~$\mu$m - $1.4$~GHz correlation. Upper right: $100$~$\mu$m - $1.4$~GHz correlation.
    Lower left: $160$~$\mu$m - $1.4$~GHz correlation. Lower right: TIR - $1.4$~GHz correlation.
    The colored and black symbols show model galaxies. For clarity, the $z \sim 0.5$ LIRGs are shown as yellow diamonds in the lower right panel.
    The solid and dotted black lines mark the model linear regression. The gray dots show the observations.
  \label{fig:galaxies_FRC_vrotDifferentForPhibbs_dwarfs_4}}
\end{figure*}

The monochromatic ($70,\ 100,\ 160$~$\mu$m) and TIR--radio correlations of all six samples are shown in Fig.~\ref{fig:galaxies_FRC_vrotDifferentForPhibbs_dwarfs_4}.
A power law of the form $L_{\rm radio} \propto L_{\rm IR}^\alpha$ is fit to the large galaxies 
using an outlier-resistant bisector fit (Kelly 2007).
The results are shown in Fig.~\ref{fig:galaxies_FRC_vrotDifferentForPhibbs_dwarfs_4}.  
We assumed uncertainties on the TIR and radio luminosities of $0.2$~dex for the local galaxies and $0.3$~dex for the 
high-z galaxies. The star-forming low-mass and dwarf galaxies nicely fall on and complement the correlations of the other galaxy samples that were
highlighted by Vollmer et al. (2025). Only the three dwarf galaxies of lowest mass show $1.4$~GHz luminosities 
higher than predicted by the monochromatic correlations.

\begin{figure*}
  \centering
  \resizebox{\hsize}{!}{\includegraphics{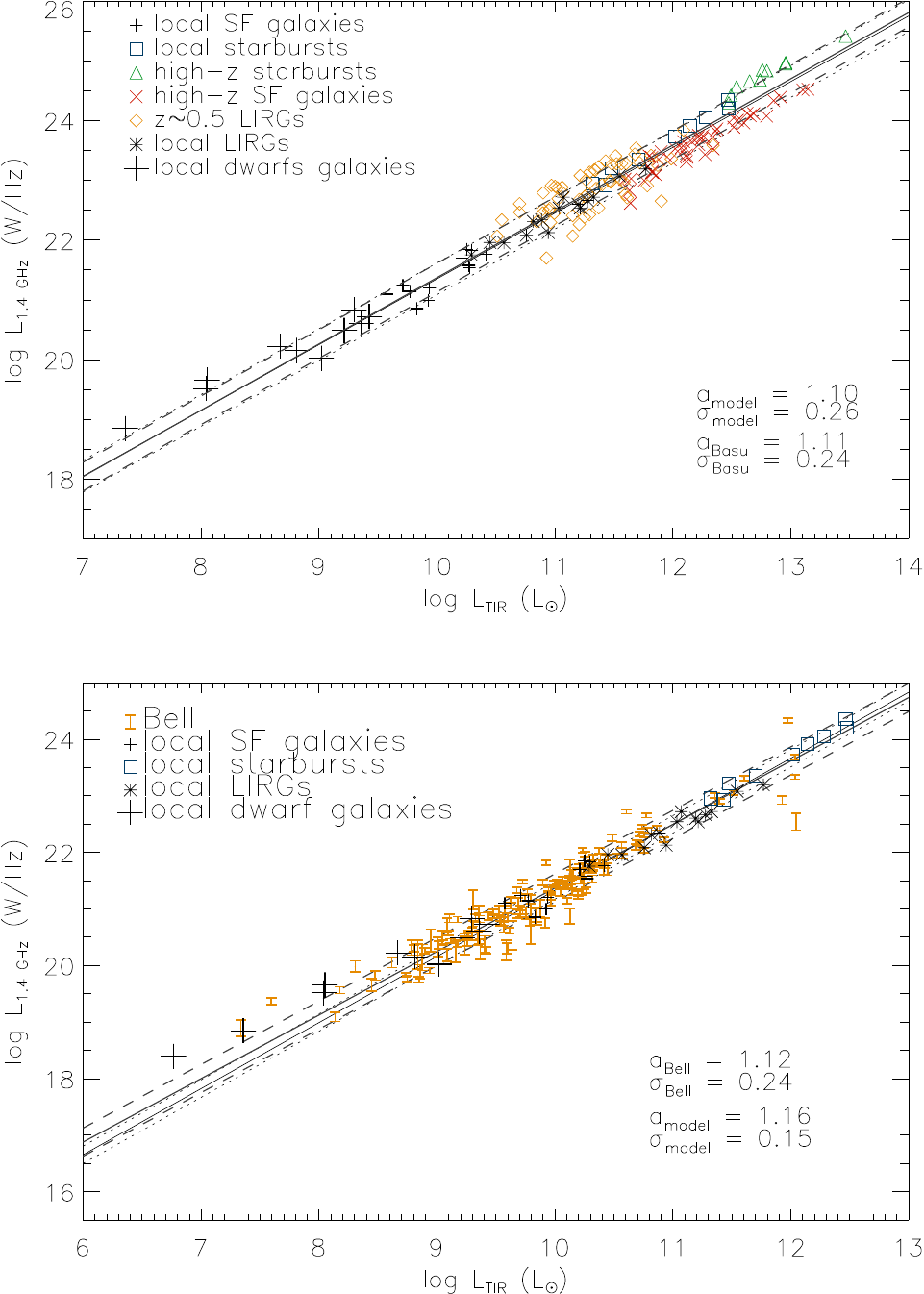}\includegraphics{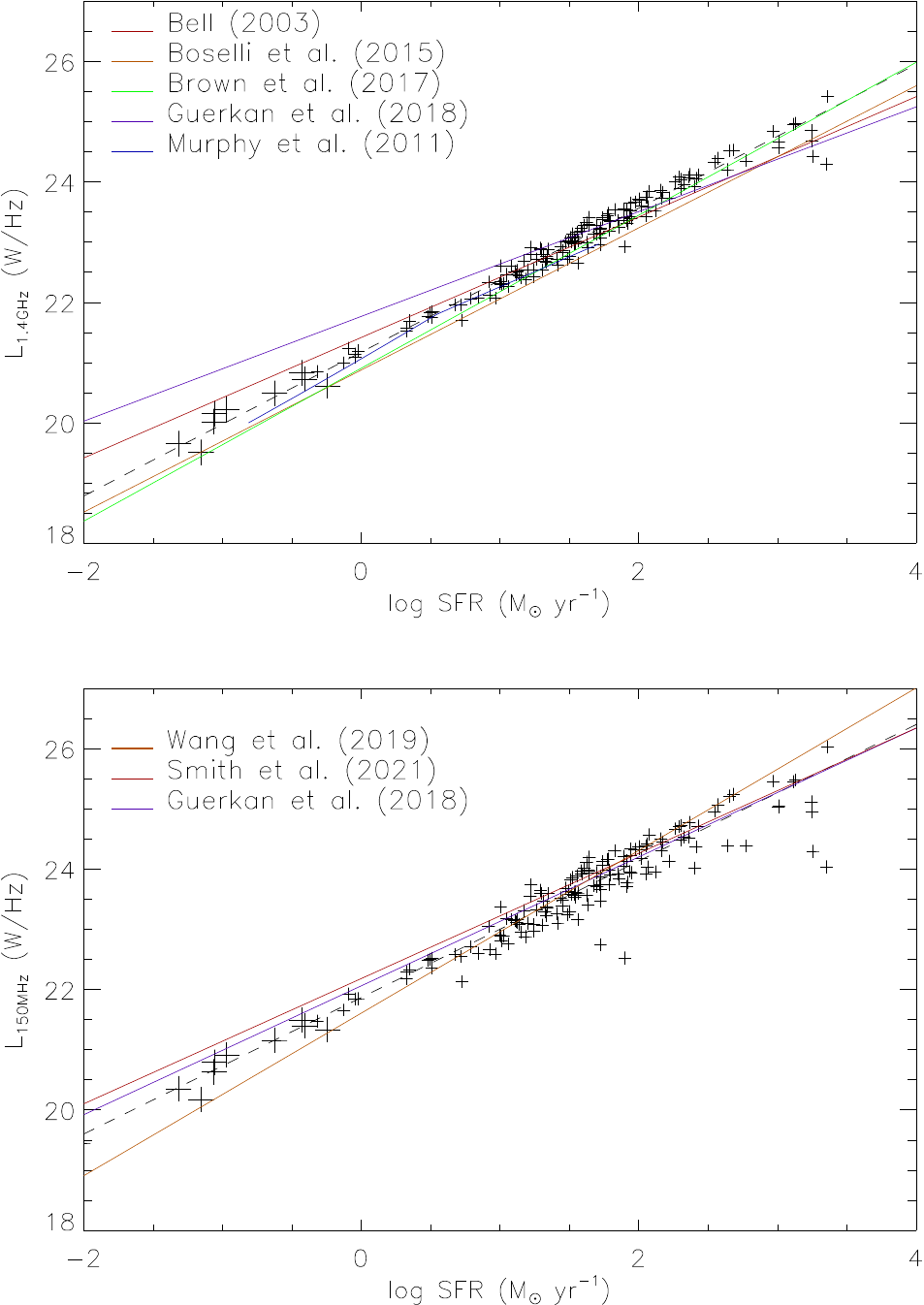}}
  \caption{Left panels: TIR--$1.4$~GHz correlations. The symbols show the model galaxies. Black solid and dotted lines show the
    model linear regression. Upper panel: Gray solid and dashed lines show the observed linear regression (Basu et al. 2015).
    Lower panel: Orange error bars show data from Bell (2003). Gray solid and dashed lines mark the observed linear regression (Bell 2003).
    Upper right panel: SFR--$1.4$~GHz correlation. Lower right panel: SFR--$150$~MHz correlation. 
    The colored lines show observed correlations of Wang et al. (2019), Smith et al. (2021), and G{\"u}rkan et al. (2018).
    The pluses mark the model galaxies. The dashed black lines in both panels correspond to outlier-resistant linear bisector fits.
  \label{fig:galaxies_FRC_vrotDifferentForPhibbs_dwarfs_5}}
\end{figure*}

Basu et al. (2015) studied the radio--TIR correlation in star-forming galaxies chosen from the PRism MUltiobject Survey up to 
redshift $1.2$ in the XMM-LSS field, employing the technique of image stacking. These authors found an exponent of the TIR--$1.4$~GHz
correlation of  $\alpha =1.11 \pm 0.04$. Bell (2003) assembled a diverse sample of local galaxies from the literature with FUV, optical, IR, and 
radio luminosities. They found a nearly linear radio--IR correlation.
For all galaxy samples except the low-mass and dwarf galaxies we calculated the slopes and offsets of the radio-TIR correlation using an outlier-resistant bisector fit
(left panels of Fig.~\ref{fig:galaxies_FRC_vrotDifferentForPhibbs_dwarfs_5}).
Our model $1.4$~GHz luminosities, and especially those of the star-forming low-mass and dwarf galaxies, are entirely consistent with the results of
Bell et al. (2003) and Basu et al. (2015).
We conclude that our sample of star-forming low-mass and dwarf galaxies follows the radio-IR correlation of the more massive star-forming galaxies.

The resulting correlations between the radio 
luminosity and the SFR are presented in the right panels of Fig.~\ref{fig:galaxies_FRC_vrotDifferentForPhibbs_dwarfs_5}
together with the measured correlations from the literature.
The fits yield $\alpha = 1.17 \pm 0.05$ at $1.4$~GHz and $\alpha = 1.12 \pm 0.05$ at $150$~MHz.

\section{Discussion \label{sec:discussion}}

Our model reproduces the IR, CO, and radio continuum observations of our sample of star-forming low-mass and dwarf galaxies within a factor of two (Sect.~\ref{sec:modelresults}).
In the following, implications on the CO conversion factor, the CO-dark gas, the thermal fraction of the radio continuum emission, the star formation law,
and the viscous gas transport will be discussed.

\subsection{The CO conversion factor \label{sec:COconv}}

\begin{figure}
  \centering
  \resizebox{\hsize}{!}{\includegraphics{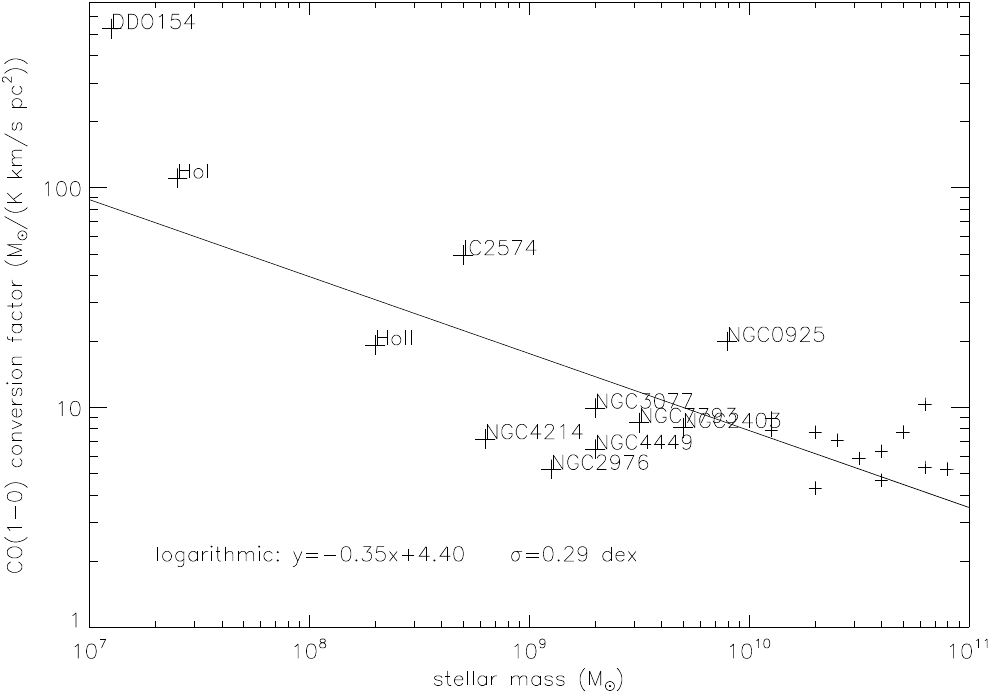}}
  \caption{CO(1-0) conversion factor as a function of stellar mass. The solid line corresponds to an outlier-robust linear regression.
  \label{fig:plots_HCNCO_dwarfs_4}}
\end{figure}

\begin{figure}
  \centering
  \resizebox{\hsize}{!}{\includegraphics{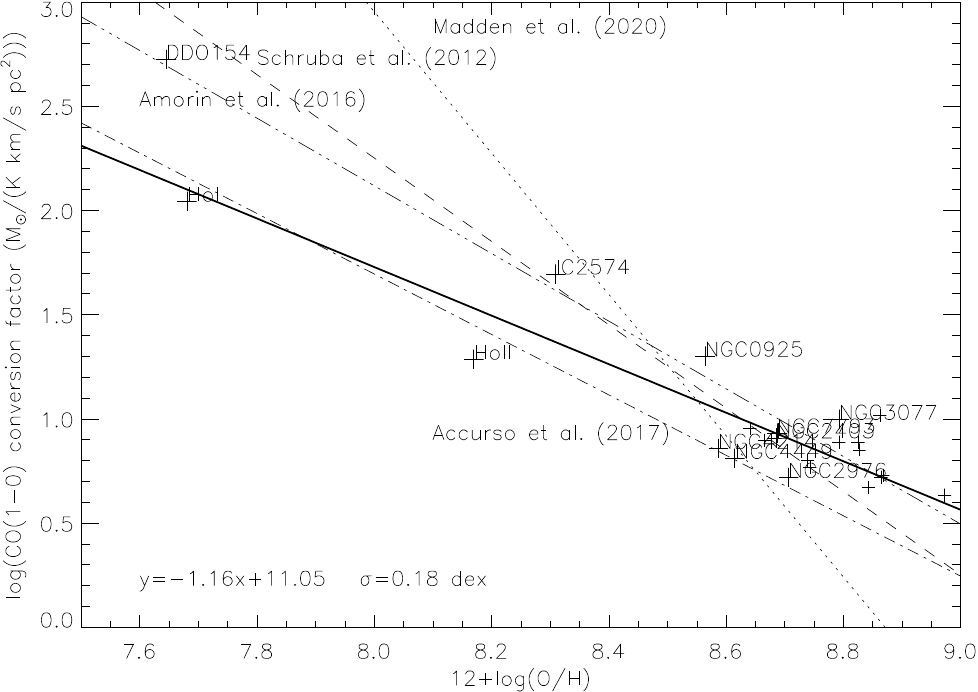}}
  \caption{Model CO(1-0) conversion factor as a function of the gas metallicity. The solid line corresponds to an outlier-robust linear regression.
    The different relations from the literature are labeled.
  \label{fig:plots_HCNCO_dwarfs_5}}
\end{figure}

Since the integrated H$_2$ gas mass and CO emission are explicitly calculated by the model, the integrated CO to H$_2$ conversion factor can be determined.
The conversion factor varies between $\sim 5$ and $\sim 500$~M$_{\odot}$(K\,km\,s$^{-1}$pc$^2$)$^{-1}$ within our sample of low-mass and dwarf galaxies
(Fig.~\ref{fig:plots_HCNCO_dwarfs_4}).
As expected, the conversion factor shows a strong dependence on stellar mass and metallicity (Fig.~\ref{fig:plots_HCNCO_dwarfs_5}).
It varies approximately linearly with metallicity: $\alpha_{\rm CO}\propto Z^{-1.2}$, very close to what Accurso et al. (2017)
 found using CO and [C{\sc ii}] observations together with multiphase numerical simulations including radiation transfer and chemical modelling.
Whereas our relation is almost inversely linear, the exponent found by Madden et al. (2020) is much steeper ($-3.4$), and those of 
Schruba et al. (2012), Amor\'{i}n et al. (2016), and Accurso et al. (2017) are somewhat steeper than our finding.
Around Solar metallicity all conversion factors, including our model, agree within a factor of two.

\begin{figure}
  \centering
  \resizebox{\hsize}{!}{\includegraphics{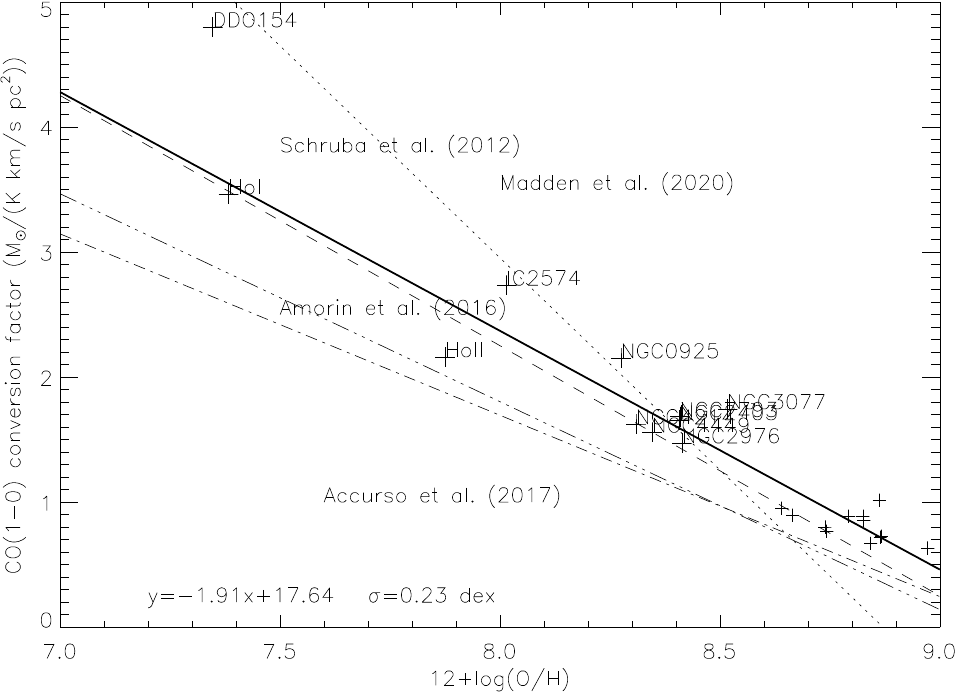}}
  \caption{Model with two times lower metallicities than shown in Fig.~\ref{fig:metallicities}.
    Model CO(1-0) conversion factor as a function of the gas metallicity. The solid line corresponds to an outlier-robust linear regression.
    The different relations from the literature are labeled.
  \label{fig:plots_HCNCO_dwarfsZ_5}}
\end{figure}

We also calculated the CO conversion factors with twice lower metallicities (Fig.~\ref{fig:plots_HCNCO_dwarfsZ_5}).
In this case, the model relation is very close to that of Schruba et al. (2012) with a slope of $-1.9$.
The CO conversion factor of DDO~154 increased by a factor of $\sim 100$ when the metallicity was divided by a factor of two, suggesting a threshold metallicity near
$12+\log({\rm O/H}=7.5$ below which virtually all of the CO is photodissociated, and the H$_2$ truly CO-dark (no reliable compensation possible via CO-H$_2$ factor).
In this $10^7$~M$_{\odot}$ dwarf galaxy the CO luminosity abruptly decreases when the metallicity is decreased by a small factor.
We do not see this kind of behavior in Ho~{\sc i}, which is only two times more massive.

\subsection{CO-dark gas \label{sec:COdark}}

We can use the model to calculate the CO luminosities without CO photodissociation (see Appendix~\ref{sec:dissociation}).
As shown in Fig.~\ref{fig:COcomparison_noCOdarkgas}, without CO photodissociation the CO luminosities are higher, 
about a factor of two higher than the detected CO luminosities.  Most galaxies with CO upper limits (except IC~2574) would have been
detected without CO photodissociation.
\begin{figure}
  \centering
  \resizebox{\hsize}{!}{\includegraphics{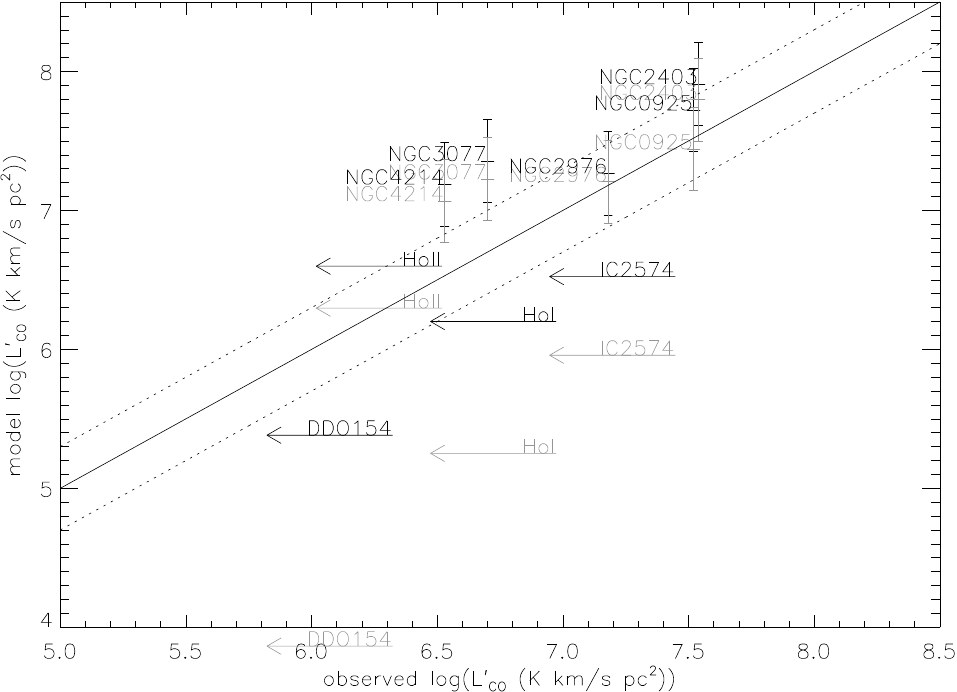}}
  \caption{Model CO luminosity without CO photodissociation as a function of the observed CO luminosity.
    The model including CO photodissociation (Fig.~\ref{fig:plots_HCNCO_dwarfs_4}) is shown in gray for comparison.
  \label{fig:COcomparison_noCOdarkgas}}
\end{figure}

\begin{figure}
  \centering
  \resizebox{\hsize}{!}{\includegraphics{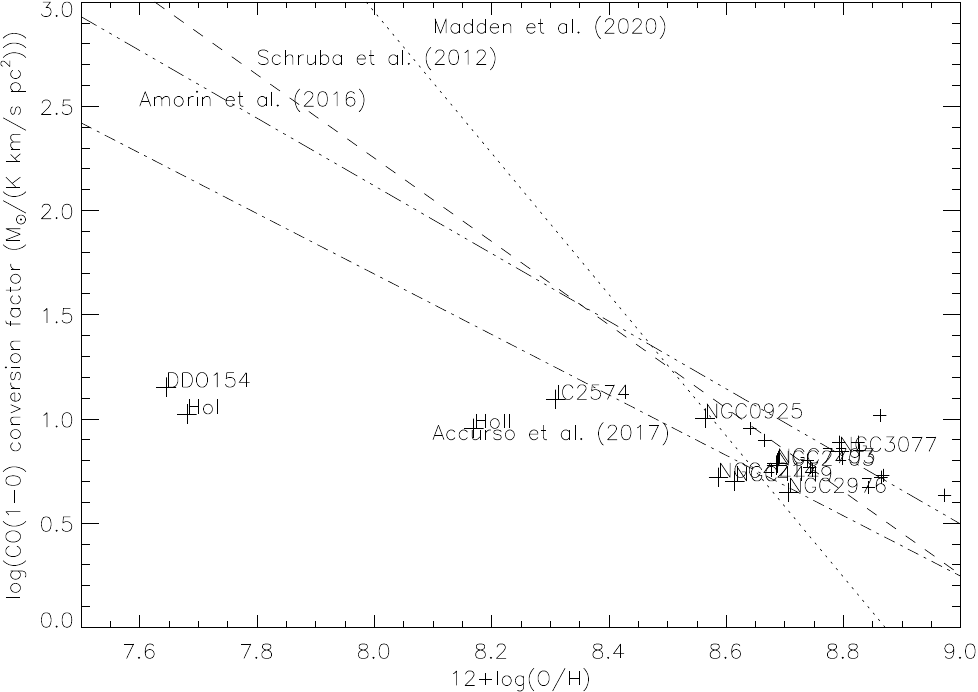}}
  \caption{CO(1-0) conversion factor as a function of  gas metallicity calculated without CO photodissociation.
  \label{fig:COcomparison_noCOdarkgas1}}
\end{figure}

The CO conversion factors without CO photodissociation are presented in Fig.~\ref{fig:COcomparison_noCOdarkgas1}.
These can be directly compared to those of Fig.~\ref{fig:plots_HCNCO_dwarfs_5}. The CO conversion factors of the
low-mass NGC galaxies are
$\alpha_{\rm CO}=8.5 \pm 2.0$~M$_{\odot}$(K\,km\,s$^{-1}$pc$^2$)$^{-1}$, a factor of two higher than the Galactic value.
Those of the four lowest mass galaxies are more than a factor of four higher than the Galactic value. 

Here we define "CO-dark gas" as being the amount of H$_2$ missed when a galactic $\ratioo$ is used to convert CO brightness to H$_2$ column density. 
The $\ratioo$ factor ($\alpha_{\rm CO}$) increases by a factor of $40$ and $10$ due to CO photodissociation in DDO~154 and Ho~{\sc i}, respectively.
It increases by a factor $4$ in IC~2574 and by a factor of two in Ho~{\sc ii} and NGC~2403. The increase is $\la 50$\,\% in the other
low-mass galaxies. This mean that $\sim 98$\,\% and $\sim 90$\,\% of the molecular hydrogen mass is CO-dark in DDO~154 and Ho~{\sc i}, respectively.
The percentage of CO-dark gas in IC~2574 is $\sim 75$\,\%, that in Ho~{\sc ii} and NGC~2403 is $\sim 50$\,\%.
Thus, only galaxies with stellar masses $\la 10^9$~M$_{\odot}$ have fractions of CO-dark gas $\ga 0.5$.

  \subsection{Intrinsic CO-H$_2$ conversion factor \label{sec:intrinsic}} 

  Running the model without CO photodissociation -- but including H$_2$ photo-dissociation -- shows that in slightly subsolar metallicity spirals (e.g. NGC~2403)
  the photodissociation of CO in the envelopes plays only a small role.
  The conversion factor $\alpha_{\rm CO}$ decreases from about 8.5~M$_{\odot}$(K\,km\,s$^{-1}$pc$^2$)$^{-1}$ (corresponding to $\ratioo=4 \times 10^{20}$~K\,km\,s$^{-1}$) to 6.0~M$_{\odot}$(K\,km\,s$^{-1}$pc$^2$)$^{-1}$.   
However, as the metallicity decreases, $\alpha_{\rm CO}$ without CO photodissociation remains essentially unchanged whereas the true $\alpha_{\rm CO}$ increases rapidly.  Photodissociation is the principal factor controlling $\alpha_{\rm CO}$ -- where H$_2$ is not photodissociated, the CO is close to optically thick in the absence of CO photodissociation. The H$_2$ part of clouds in galaxies is intrinsically similar despite the highly different conditions.

In spiral disks 
the UV brightness decreases by more than an order of magnitude between the central regions and $R_{25}$ whereas the metallicity decreases by a factor $\sim2$.  Thus, the weak CO emission in outer disks, beyond the effective radius, of solar or slightly subsolar spiral galaxies is not due to CO photodissociation but to the lower excitation and lower molecular fraction.

\subsection{The thermal fraction of the radio continuum emission \label{sec:thermalf}}

The radio continuum SEDs of the low-mass and dwarf galaxies (Fig.~\ref{fig:radiospectra_dwarfs}) become flatter towards high frequencies
because of an increasing fraction of emission of thermal electrons (bremsstrahlung). We calculated the thermal fraction for the different galaxy samples
at frequencies between $150$~MHz and $30$~GHz (Table~\ref{tab:thermalf}). As expected, the thermal fraction increases from a few percent at $150$~MHz
to $50$--$70$\,\% at $30$~GHz. Interestingly, the highest thermal fractions are observed in our star-forming low-mass and dwarf galaxy sample.
At $1.4$~GHz, the mean thermal fraction is $23$\,\%. DDO~154 and IC~2574 have a thermal fraction of $45$\,\%, which is consistent with the results of Roychowdhury \& Chengalur (2012)
who found a thermal fraction of $\sim 50$\,\% for their sample of
faint dwarf irregular galaxies with typical SFRs of several $10^{-3}$~M$_{\odot}$yr$^{-1}$.
For our low-mass and dwarf galaxy sample the thermal fraction is $40$\,\% at $5$~GHz. It increases to $70$\,\% at $30$~GHz. 
Presumably this is due to the weaker large-scale magnetic field in the smaller galaxies (see also Klein et al. 1984).
\begin{table*}
      \caption{Thermal fraction in percent.}
         \label{tab:thermalf}
      \[
       \begin{tabular}{rrrrrrrr}
         \hline
         & local & local & local  & high-z & z=0.5   & z=1  &  local  \\
         & dwarfs & MS & SB & SB & MS  & MS &  LIRGs \\
         \hline
   150~MHz  &      6 &            3 &           13 &           10 &            3 &            2 &            3 \\
   325~MHz   &     10 &            5 &            9 &            7 &            4 &            3 &            5 \\
   610~MHz  &     15 &            7 &            8 &            6 &            5 &            3 &            7 \\
   1.4~GHz   &     23 &           12 &            8 &            7 &            8 &            6 &           10 \\
   5~GHz  &     41 &           26 &           13 &           13 &           17 &           12 &           20 \\
   10~GHz   &     53 &           38 &           18 &           20 &           25 &           20 &           30 \\
   30~GHz  &     72 &           60 &           34 &           38 &           45 &           38 &           50 \\
        \hline
        \end{tabular}
      \]
\end{table*}

\subsection{The star formation law \label{sec:SFRlaw}}

The star formation efficiency  
in local main sequence galaxies is roughly linear with a gas depletion time of
$SFE = M(H_2)/SFR \sim2$ Gyr (e.g., Bigiel 2008, 2011). Saintonge et al. (2011a) 
compared the molecular gas depletion time and global galaxy parameters in
a sample of $\sim 300$ local galaxies with $10^{10} < M_* < 10^{11.5}$~M$_{\odot}$.
The global molecular gas depletion time was found to vary with stellar mass, stellar surface
mass density, concentration of the light, near-ultraviolet (NUV)-r color, and speciﬁc star formation rate (sSFR). The strongest dependencies were on color and sSFR.

\begin{figure}
  \centering
  \resizebox{\hsize}{!}{\includegraphics{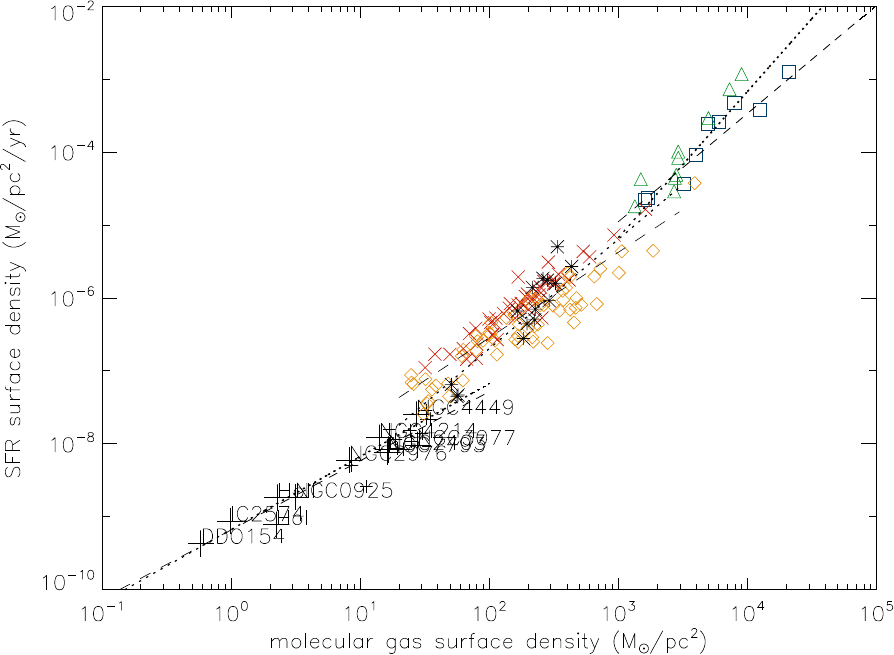}}
  \caption{Star formation rate surface density as a function of the molecular gas surface density. The symbols are the same as in Fig.~\ref{fig:plots_HCNCO_dwarfs_1}.
    The dashed lines correspond to outlier-robust linear regressions
    to three sub-samples (local main sequence, intermediate and high-z main sequence, local and high-z starbursts). The dotted lines correspond to exponents of
    the logarithmic relation of $1$, $1.5$, and $2$, respectively.
  \label{fig:plots_1_phibss2_dwarfs10}}
\end{figure}

The link between model $\dot{\Sigma}_*$ (SF surface density) and model $\Sigma(H_2)$ (H$_2$ surface density) is presented in Fig.~\ref{fig:plots_1_phibss2_dwarfs10}.
The slope steepens with increasing molecular gas surface density, increasing from linear in 
local main sequence galaxies to $\sim 1.5$ for local LIRGs and $0.5 \le z \le 1.5$ main sequence galaxies to $\sim 2$ for local and high-z starbursts.
Part of the steepening could be because at high pressure all of the gas is molecular, increasing the SFR.  At high redshift, the galaxies are physically small so the surface densities are high.

\begin{figure}
  \centering
  \resizebox{\hsize}{!}{\includegraphics{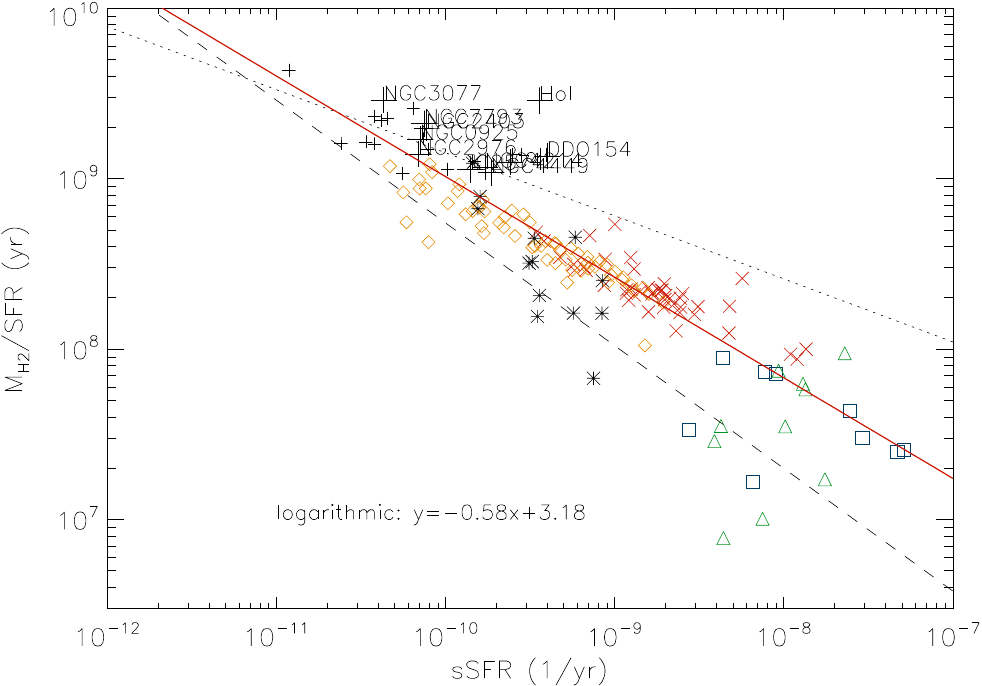}}
  \caption{Molecular gas depletion time as a function of the sSFR. The red solid line corresponds to an outlier-robust linear regression to our model galaxies.
     The symbols are the same as in Fig.~\ref{fig:plots_HCNCO_dwarfs_1}.
    The dotted line corresponds to the relation found by Huang \& Kauffmann (2014), the dashed line to that of Saintonge et al. (2011b).
  \label{fig:plots_HCNCO_dwarfs_36}}
\end{figure}

The molecular gas depletion time as a function of the sSFR is shown in Fig.~\ref{fig:plots_HCNCO_dwarfs_36} together with the relations determined by
Saintonge et al. (2011b) and Huang \& Kauffmann (2014). 
$sSFR =\dot{M_*}/M_*$ and $t_{\rm dep} = M(H_2)/SFR$ so the relation (red line) shown is 
$t_{\rm dep} \approx 10^{3.18} sSFR^{-0.58}$ .    
The scatter is $0.12$~dex, about half that of Saintonge et al. (2011b) or Huang \& Kauffmann (2014). The relation found by Saintonge et al. (2011b)
is somewhat steeper (exponent of $-0.72$), that of Huang \& Kauffmann (2014) is shallower ($-0.37$) than our model relation.
All galaxies except Ho~{\sc i} lie on the model relation. The galaxy sample with the largest scatter is that of the high-z starburst galaxies.
It should be noted that the variation in sSFR is very small among spirals (little variation around sSFR$\approx 5 \times 10^{-11}$ and the phase during which a galaxy (necessarily starburst) has a high sSFR is short.  Small galaxies, generally chemically young, have much smaller stellar masses and, unless they have little or no gas (like local dwarf spheroidals), have higher sSFR and lower gas consumption times.

Since the model SF surface density is proportional to the total gas pressure (Fig.~A.1 of Vollmer et al. 2025, Ostriker \& Kim 2022),
a change of regime from vertically non-selfgravitating to self-gravitating gas might lead to the steepening of the SFR correlation with increasing gas surface density.
To investigate the influence of vertical self-gravity ($0.5\,\pi\,G\,\Sigma^2$; Eq.~\ref{eq:pressure}) on the star formation law, we calculated the ratio between the pressure
of self-gravitating gas to the total gas pressure:
\begin{equation}
  f_{\rm selfgrav}=\int \left(\rho_{\rm gas}\,v_{\rm turb}^2/(\pi G \Sigma^2_{\rm gas})\right)\dot{\Sigma}_* R\,dR/\int \dot{\Sigma}_* R\,dR\ .
\end{equation}
Since this ratio does not vary significantly between the main-sequence and starburst galaxies and since $f_{\rm selfgrav}$ for
local and high-z starburst galaxies are significantly different (Table~\ref{tab:selfgrav}), we conclude that vertical self-gravity does
not play an important role in shaping the $\dot{\Sigma}_*$--$\Sigma_{\rm H_2}$ relation.

We also calculated our model without vertical gas self-gravity. This led to much higher gas velocity dispersions, approaching the rotation velocities,
compared to the models with self-gravity. The resulting logarithmic $\dot{\Sigma}_*$--$\Sigma_{\rm H_2}$ and SFR--sSFR relations have similar slopes
as those of the model with self-gravity, corroborating our conclusion that self-gravity is not responsible for the different slopes of the
$\dot{\Sigma}_*$--$\Sigma_{\rm H_2}$ relation.
We argue that it is the combination of a constant mass accretion rate, a star formation surface density that is proportional to the gas pressure,
a constant Toomre $Q$ parameter, and energy flux conservation (Appendix~\ref{sec:dgasdisk}) that lead to the observed $\dot{\Sigma}_*$--$\Sigma_{\rm H_2}$ and SFR--sSFR relations.
\begin{table}
      \caption{Self-gravitating gas disk pressure fraction.}
         \label{tab:selfgrav}
      \[
       \begin{tabular}{lc}
        \hline
        sample & $f_{\rm selfgrav}$ \\
        \hline
        local low-mass and dwarf galaxies & $0.20 \pm 0.02$ \\
        local main sequence galaxies & $0.29 \pm 0.07$ \\
        local starburst galaxies & $0.45 \pm 0.09$ \\
        high-z starburst galaxies & $0.33 \pm 0.14$ \\
        z=0.5 main sequence galaxies & $0.48 \pm 0.08$ \\
        z=1 main sequence galaxies & $0.52 \pm 0.07$ \\
        local Luminous InfraRed Galaxies & $0.56 \pm 0.08$ \\
        \hline
        \end{tabular}
      \]
\end{table}

\subsection{Radial gas transport \label{sec:radialgt}}

\begin{figure}
  \centering
  \resizebox{\hsize}{!}{\includegraphics{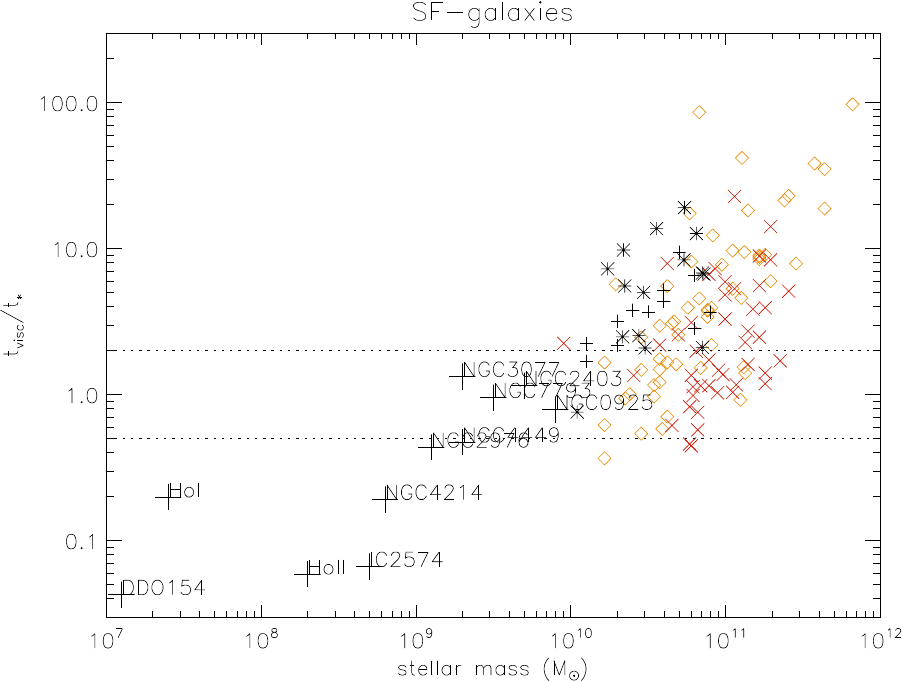}}
  \caption{Fraction $t_{\rm visc}/t_*$ as a function of the stellar mass of the model galaxies. Upper panel: SF galaxies.
    The $t_{\rm visc}/t_*$ fractions of the star-forming galaxies should thus be taken as lower limits
    with an uncertainty of a factor of two towards higher values. The symbols are the same as in Fig.~\ref{fig:plots_HCNCO_dwarfs_1}.
  \label{fig:plots_1_phibss2_dwarfs14}}
\end{figure}

The turbulent viscosity of the gas is given by $\nu=v_{\rm turb}\,l_{\rm driv}$. To compare the radial gas accretion rate with 
gas consumption by star formation, we evaluate the star formation timescale ($t_*=\Sigma/\dot{\Sigma}_*$) and the viscous
timescale ($t_{\rm visc}=(R^2\Omega)'/(-\Omega' \nu$)) at the effective (half-light) radius of each galaxy.
Fig.~\ref{fig:plots_1_phibss2_dwarfs14} shows the model $t_{\rm visc}/t_*$ ratio versus stellar mass for the star-forming galaxies.
We found $t_{\rm visc}/t_* \la 1$ for all low-mass and dwarf galaxies.
Therefore, as argued in Vollmer et al. (2025), viscous gas transport plays a role in star-forming galaxies with $M_* \la 10^{10}$~M$_{\odot}$.

\section{Conclusions \label{sec:conclusions}}

We extended the work of Vollmer et al. (2025) on local and high-z star-forming and starburst galaxies by adding local star-forming low-mass and dwarf galaxies.
The gas chemistry computed by the Nautilus code depends on the cosmic ray (CR) ionization rate, which was calculated following the analytical model of Pohl (1993)
using the local density, magnetic field strength, and radiation field of our analytical model. The injection rate of CR particles is proportional
to the SFR per volume ($\dot{\rho}_*$). The molecular line emission is calculated by RADEX.
We then calculated the radio continuum emission of the galaxies following the framework of Vollmer et al. (2022, 2025)
and Condon \& Ransom (2016) and Pacholczyk (1970).
For the CR energy loss terms we used the formalism of Pohl (1993) with the energy loss time scales of Vollmer et al. (2022).
The new method has the advantage that the normalization of the CRe density and, as a consequence, of the CR ionization rate is explicitly set
within the model framework. The normalization of the CR ionization rate we found for the different galaxy samples 
is higher by a factor of three than the normalization 
in the solar neighborhood. This means that the mean yield of low-energy CR particles for a given SF surface density is
higher by three times in external galaxies than was observed by Voyager~I.
We found that the thermal fraction of the radio continuum emission of our star-forming low-mass and dwarf galaxy sample is higher than $40$\,\% for frequencies
higher than $5$~GHz.

The model was applied to $7$ samples of main-sequence and starburst galaxies (see Sect.~\ref{sec:observations}) at low, intermediate, and high redshifts. Based on the 
comparison between the model and available observations we conclude that
\begin{enumerate}
\item
  The model reproduces the IR (Sect.~\ref{sec:IR}), CO (Sect.~\ref{sec:CO}), and radio continuum (Sect.~\ref{sec:RC}) luminosities to within a factor of two.
\item
  The model CO-H$_2$ conversion factor $\alpha_{CO}$ varies with metallicity as $\alpha_{\rm CO(1-0)} \propto 1/Z$ (Sect.~\ref{sec:COconv}).
\item
   The model $\alpha_{\rm CO(1-0)}$ varies with stellar mass as $M_*^{1/3}$ (Fig.~\ref{fig:plots_HCNCO_dwarfs_4}). 
\item
  Photodissociation of CO in cloud envelopes ("CO-dark gas") is responsible for the variation of of $\alpha_{\rm CO}$ with $M_*$ and $Z$.  Without CO photodissociation, $\alpha_{\rm CO}$ is almost constant (Sect.~\ref{sec:COdark} and \ref{sec:intrinsic}).
\item
  Star-forming low-mass and dwarf galaxies follow the radio--IR, and radio--SFR correlations (Sect.~\ref{sec:RC}).
\item
  The star formation rate vs H$_2$ mass (expressed as surface densities) is a three-part function highly dependent on the subsample (Sect.~\ref{sec:SFRlaw}).
\item
  Viscous gas transport is important in galaxies with $t_{\rm visc} \la t_{\rm depl}$, which generally corresponds to  $M_* \la 10^{10}$~M$_{\odot}$ (Sect.~\ref{sec:radialgt}).
\end{enumerate}


\FloatBarrier

\begin{appendix}

  \onecolumn
  
\section{Gas metallicity\label{sec:gmetal}}

\begin{figure}[h!]
  \centering
  \resizebox{8cm}{!}{\includegraphics{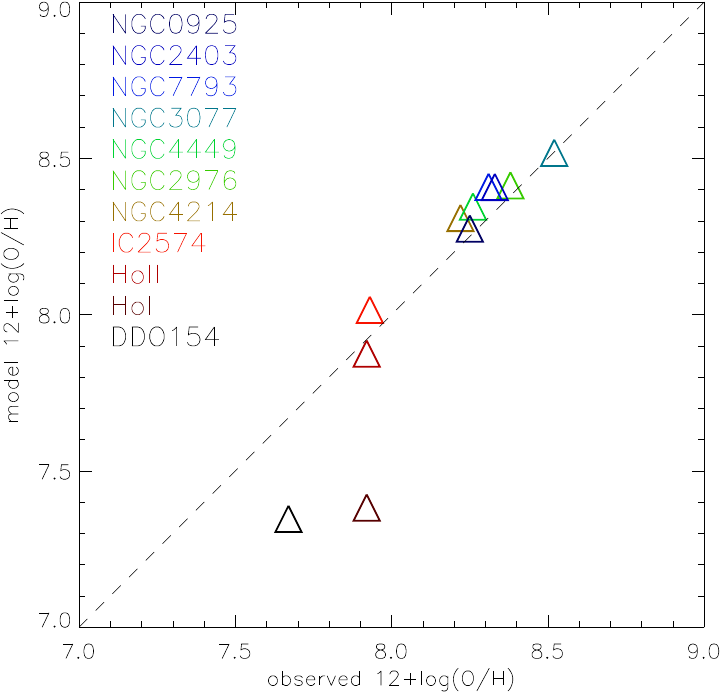}}
  \caption{Model gas metallicity as a function of the observed gas metallicity. The dashed line corresponds to the one-to-one relation.
  \label{fig:metallicities}}
\end{figure}
\FloatBarrier

\section{IR SEDs \label{sec:irseds}}

\begin{figure*}
  \centering
  \resizebox{14cm}{!}{\includegraphics{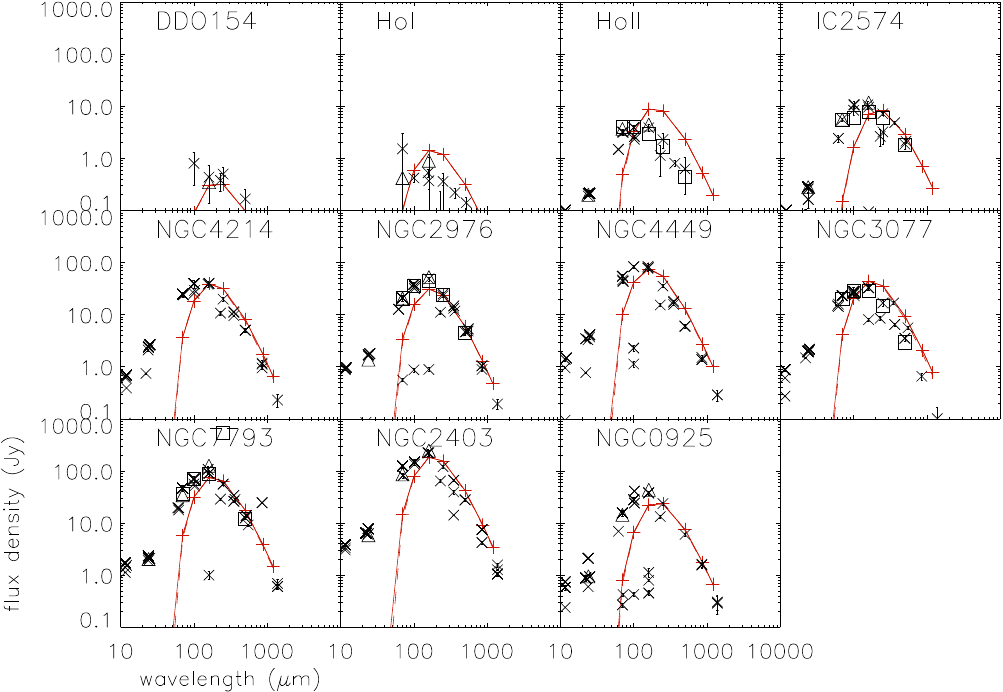}}
  \caption{Infrared spectral energy distributions. Red pluses and solid line: model SED.
    Red dashed line: modified Planck fit for temperature determination. 
    Black crosses: VizieR photometry. Black boxes: Herschel data from Dale et al. (2012). The errors bars are shown, if present in the VizieR tables,
    but often barely visible.
  \label{fig:IRspectra_dwarfsZ}}
\end{figure*}
\FloatBarrier

\section{Radio SEDs \label{sec:radioseds}}

  \begin{figure*}[h!]
  \centering
  \resizebox{14cm}{!}{\includegraphics{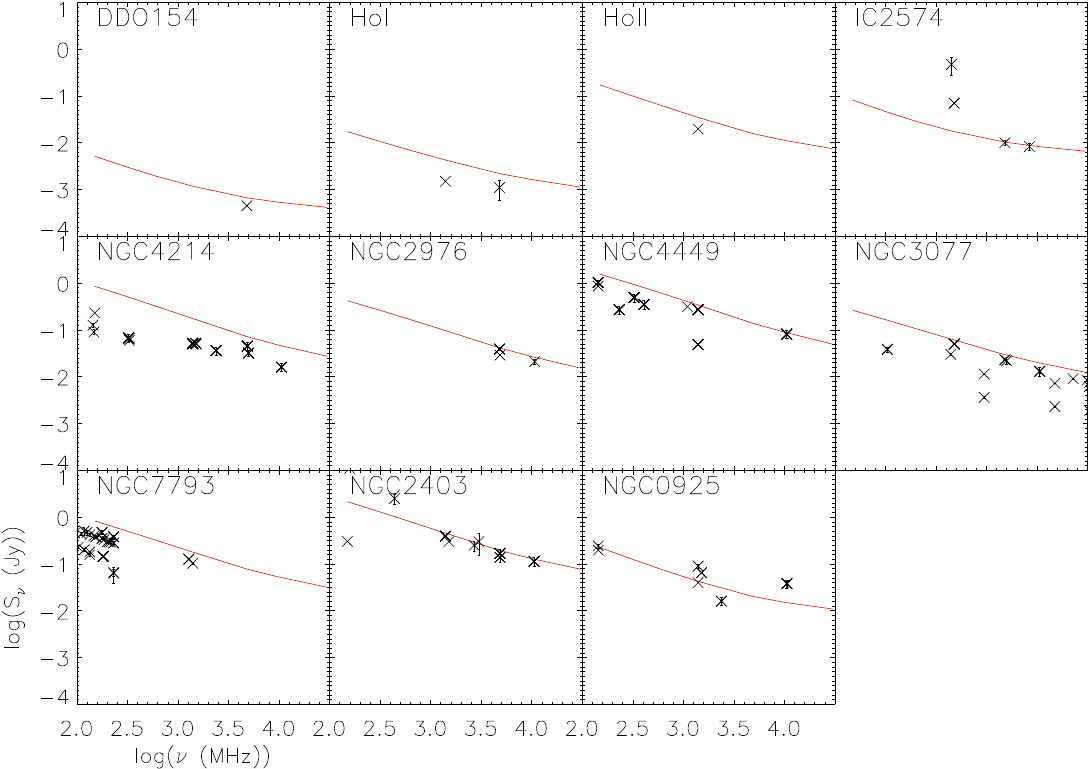}}
  \caption{Radio continuum spectral energy distributions. The model results are shown by red lines. Black crosses: VizieR photometry.
    The errors bars are shown, if present in the VizieR tables, but often barely visible.
  \label{fig:radiospectra_dwarfs}}
\end{figure*}
\FloatBarrier

\twocolumn
  
\section{The model gas disk \label{sec:dgasdisk}}

In this section and the following, we briefly describe the Vollmer et al. (2017)  model with the minor modifications given in Vollmer et al. (2025).
The meaning of the different variables is given in Appendix~\href{https://doi.org/10.5281/zenodo.14264985}{C}.
Following Vollmer \& Leroy (2011) and Vollmer et al. (2021), the SFR per unit volume is given by
\begin{equation}\label{eq:star-form}
\dot{\rho}_{*} = \Phi_{\rm V} \frac{\rho}{t_{\rm ff}^{\rm l}}\ .
\end{equation}
where $\Phi_{\rm V}$ is the volume filling factor, $\rho$ is the large-scale density, $l$ and $\rho_{\rm l}=\rho/\Phi_{\rm V}$ the size and density of selfgravitating clouds,
and $t_{\rm ff}^{\rm l}=\sqrt{3\pi/(32\,G\,\rho_{\rm l})}$ the local free-fall time.
This prescription leads to an SFR per unit area which is proportional to the gas pressure $p=\rho v_{\rm turb}^2$ (Fig.~A.1 of Vollmer et al. 2022)
as expected if star formation is pressure-regulated and feedback-modulated (Ostriker \& Kim 2022).

For self-gravitating clouds with a virial parameter of two (Bertoldi \& McKee 1992), the turbulent crossing time $t_{\rm turb,cl}$ equals
$1.4$ times the free-fall time
$t_{\rm turb,cl}$:
\begin{equation}
t_{\rm turb,cl} =  \frac{l_{\rm cl}}{2\,v_{\rm turb,cl}} = 1.4 \times t_{\rm ff,cl}=1.4 \times \sqrt{\frac{3 \pi \Phi_{\rm V}}{32 G \rho}}
\label{eq:tturbcl}
\end{equation}
where $l_{\rm cl}$ and $v_{\rm turb,cl}$ are the size and the turbulent 1D velocity dispersion of a single gas cloud, respectively. Following Larson's
law (Larson 1981), we can simplify the expression of the turbulent crossing time:
\begin{equation}
t_{\rm turb,cl}=\frac{l_{\rm cl}}{2\,v_{\rm turb,cl}} = \frac{l_{\rm driv}}{2\,v_{\rm turb} \sqrt{\delta}}\ .
\end{equation}
where $\delta$ is the scaling between the driving length scale and the
size of the largest self-gravitating structures, such as $\delta = l_{\rm driv}/l_{\rm cl}$.
The expression for the SFR then becomes
\begin{equation}
  \label{eq:sfrrec}
\dot{\rho}_*= 2.8\,\sqrt{\delta} \Phi_{\rm V} \rho \frac{v_{\rm turb}}{l_{\rm driv}}\ .
\end{equation}
and $\dot{\Sigma}_*=\dot{\rho}_* l_{\rm driv}$.

The SFR density is used to calculate the rate of energy injection by supernovae (SNe):
\begin{equation}
  \label{eq:energyflux}
  \frac{\dot{E}_{\rm SN}}{\Delta A}=\xi\,\dot{\Sigma}_{*} 
  = \xi\,\dot{\rho}_{*} l_{\rm driv}=\frac{1}{2} \Sigma \nu \frac{v_{\rm turb,3D}^{2}}{l_{\rm driv}^{2}}
  = \frac{3\,\sqrt{3}}{2} \Sigma \frac{v_{\rm turb}^3}{l_{\rm driv}}\ ,
\end{equation}
where $\Delta A$ is the unit surface element of the disk, $\dot{\Sigma}_{*}$ and $\dot{\rho}_{*}$ the star formation per unit area and unit volume,
$\Sigma$ the total gas surface density, $\nu$ the turbulent viscosity, $v_{\rm turb}$ the turbulent velocity of the gas, and the CO disk thickness
is assumed to be $l_{\rm driv}$. The turbulent viscosity is $\nu=\sqrt{3}v_{\rm turb} l_{\rm driv}$.
The energy injection rate is related to the turbulent
velocity dispersion and the driving scale of turbulence. These quantities in turn provide estimates of the clumpiness of gas in the disk (i.e., the
contrast between local and average density) and the rate at which viscosity moves matter inward.
The derived expressions for $\Phi_{\rm V}$ and $\dot{\Sigma}_*$ are
\begin{equation}
\Phi_{\rm V}=\frac{4.1\,G\,\rho\,l_{\rm driv}^2}{3\,\pi\,\delta\,v_{\rm turb}^2}
\end{equation}
and
\begin{equation}
\dot{\Sigma}_*=\frac{11.4\,G\,\rho^2 l_{\rm driv}^2}{3\,\pi\,\sqrt{\delta} v_{\rm turb}}\ .
\label{eq:sigmastar1}
\end{equation}

In the model, the disk scale height is
determined unambiguously by the assumption of hydrostatic equilibrium and the turbulent pressure
(Elmegreen 1989):

\begin{equation}
p_{\rm turb}=\rho v_{\rm  turb}^{2} = \frac{\pi}{2} G \Sigma ( \Sigma + \Sigma_{*} \frac{v_{\rm turb}}{v_{\rm disp}^{*}})~,
\label{eq:pressure}
\end{equation}

\noindent where $v_{\rm disp}^{*}$ is the stellar vertical velocity dispersion 
and $\Sigma$ the surface density of gas and stars. The stellar velocity dispersion is calculated
by $v_{\rm disp}^{*}=\sqrt{2 \pi G \Sigma_{*} H_{*}}$, where
the stellar vertical height is taken to be $H_{*}=l_{*}/7.3$ with  $l_{*}$ being the
stellar radial scale length (Kregel et al. 2002). 
We neglect thermal, cosmic ray, and magnetic pressure. 

The model relies on several empirical calibrations: the relation between the stellar
velocity dispersion and the stellar disk scale length, the relationship between the star-formation rate (SFR) and the energy injected into
the ISM by SNe, and the characteristic time of H$_2$ formation, which is related to the gas metallicity and the gas density.
The fraction between the molecular and the total gas mass is governed by the turbulent crossing time $t^{\rm l}_{\rm turb}$ and the molecule formation timescale.
In addition, photodissociation of molecules is taken into account.

The model is divided into a large-scale and a small-scale part. The former yields the surface density, turbulent velocity, disk
height, and gas viscosity. The latter is relevant for self-gravitating gas clouds $(t^{\rm l}_{\rm ff} \le t^{\rm l}_{\rm turb})$.
The non-self-gravitating and self-gravitating clouds obey different scaling relations, which are set by observations.
For each gas density, the mass fraction of clouds of this density is characterized by a log-normal probability distribution function and the Mach number
(Padoan et al. 1997). The temperatures of the gas clouds are calculated via the equilibrium between turbulent mechanical and
cosmic ray heating and gas cooling via CO and H$_2$ line emission. The abundances of the different molecules are determined using the time-dependent
gas-grain code Nautilus (Hersant et al. 2009, Ruaud et al. 2016). The dust temperatures are calculated via the equilibrium between heating by the interstellar UV and optical
radiation field and cooling via IR emission.

\section{Molecular fraction based on photodissociation \label{sec:dissociation}}

For the determination of the H$_2$ column density of a gas cloud, we take into account 
(i) photo-dissociation of H$_2$ molecules and (ii) the influence of the finite cloud lifetime on the H$_2$ formation.

For the photo-dissociation of H$_2$ molecules, we follow the approach of Krumholz et al. (2008, 2009). 
These authors solved the idealized problem of determining the location of the atomic-to-molecular transition in a uniform spherical 
gas cloud bathed in a uniform isotropic dissociating radiation field. It is assumed that the transition from atomic to molecular gas 
occurs in an infinitely thin shell. The cloud has a constant inner molecular and outer atomic gas density.
The inner molecular core and the outer atomic shell are assumed to be in thermal pressure equilibrium.
The H$_2$ to H{\sc i} ratio is 
\begin{equation}
R_{\rm H_2} \simeq \big(1+(s/11)^3(\frac{125+s}{96+s})^3)\big)^{\frac{1}{3}}-1
,\end{equation}
with $s=(\Sigma_{\rm cl}/1~{\rm M}_{\odot})(Z/Z_{\odot})/(4\,\tau_{\rm H{\sc i}})$. The H{\sc i} optical depth is
\begin{equation}
\tau_{\rm H{\sc I}}=\frac{\chi}{4} \frac{2.5+\chi}{2.5+\chi {\rm e}}\ ,
\end{equation}
with the dimensionless radiation field strength $\chi$ , which we set to
$\chi=3.1\,(\dot{\Sigma_*}/(10^{-8}~{\rm M}_{\odot}{\rm pc}^{-2}{\rm yr}^{-1}))/(n_{\rm cl}/(100~{\rm cm}^{-3}))$.
Here, we assume a constant ratio between the inner molecular and outer atomic gas density, which is of the order of $10$.
For $\tau_{\rm H{\sc I}}=\frac{1}{4}$ and solar metallicity, the transition between a molecular- and atomic-dominated cloud occurs
at $\Sigma_{\rm cl} \simeq 20$~M$_{\odot}$pc$^{-2}$.
The H$_2$ fraction of the cloud is $f_{\rm H_2}=R_{\rm H_2}/(1+R_{\rm H_2})$.
This treatment insures a proper separation of H{\sc i} and H$_2$ in spiral galaxies, that is, 
clouds of low density ($\sim 100$~cm$^{-3}$) and low column density 
($\sim 10^{21}$~cm$^-2$) are fully atomic, whereas clouds of high density, that is, GMCs, ($\geq 1000$~cm$^{-3}$) and high column 
density ($\geq 10^{22}$~cm$^-2$) are fully molecular. In starburst regions (e.g., in ULIRGs), where gas densities and surface 
densities are much higher, this treatment has no effect, since the gas will be fully molecular.

In a second step, we take into account the molecular fraction due to the finite lifetime of the gas cloud 
$f_{\rm mol}^{\rm life}=t_{\rm ff}^{\rm cl}/t{\rm mol}^{\rm cl}/(1+t_{\rm ff}^{\rm cl}/t_{\rm mol}^{\rm cl})$.
The total molecular fraction of a cloud is $f_{\rm mol}=f_{\rm mol}^{\rm life} \times f_{\rm mol}^{\rm diss}$. The molecular fraction
due to the finite lifetime $f_{\rm mol}^{\rm life}$ has the highest influence on $f_{\rm mol}$ at large galactic radii.

We now go from the H$_2$ mass fraction to the CO mass fraction.
In an externally irradiated gas cloud, a significant H$_2$ mass may lie outside
the CO region, that is, it is dark in the outer regions of the cloud where the gas phase carbon resides in C or C$^+$. 
In this region, H$_2$ self-shields or is shielded by dust from UV photodissociation, whereas CO is photodissociated.
Following Wolfire et al. (2010), the dark gas mass fraction for a cloud of constant density is 
\begin{equation}
\label{eq:fdg}
f_{\rm DG}=\frac{M_{\rm H_2}-M_{\rm CO}}{M_{\rm H_2}}=1-\big(1 - \frac{2 \Delta A_{\rm V, DG}}{A_{\rm V}}\big)^3
,\end{equation}
with
\begin{equation}
\begin{split}
\Delta A_{\rm V,DG}=&0.53-0.045\,{\rm ln}\big(\frac{\dot{\Sigma_*}/(10^{-8}~{\rm M}_{\odot}{\rm pc}^{-2}{\rm yr}^{-1})}{n_{\rm cl}}\big)-\\
&0.097\,{\rm ln}\big(\frac{Z}{Z_{\odot}}\big)
\end{split}
,\end{equation}
and $A_{\rm V}=2\,(Z/Z_{\odot})N_{\rm cl}/(1.9 \times 10^{21}~{\rm cm}^{-2})$ where $N_{\rm cl}$ is the H$_2$ column density.
The CO mass fraction is then $f_{\rm CO}=f_{\rm H_2} \, \big(1 - \frac{2 \Delta A_{\rm V, DG}}{A_{\rm V}}\big)^3$.

\end{appendix}




\begin{thebibliography}{}

\bibitem[Accurso et al.(2017)]{2017MNRAS.470.4750A} Accurso, G., Saintonge, A., Catinella, B., et al.\ 2017, \mnras, 470, 4, 4750
  
\bibitem[Amor{\'\i}n et al.(2016)]{2016A&A...588A..23A} Amor{\'\i}n, R., Mu{\~n}oz-Tu{\~n}{\'o}n, C., Aguerri, J.~A.~L., et al.\ 2016, \aap, 588, A23

\bibitem[Aniano et al.(2020)]{2020ApJ...889..150A} Aniano, G., Draine, B.~T., Hunt, L.~K., et al.\ 2020, \apj, 889, 2, 150
  
\bibitem[Basu et al.(2015)]{2015ApJ...803...51B} Basu, A., Wadadekar, Y., Beelen, A., et al.\ 2015, ApJ, 803, 51

\bibitem[Beck(2015)]{2015A&ARv..24....4B} Beck, R.\ 2015, \aapr, 24, 4
  
\bibitem[Bell(2003)]{2003ApJ...586..794B} Bell, E.~F.\ 2003, ApJ, 586, 794 

\bibitem[Benson et al.(2013)]{2013ApJ...770...76B} Benson, A., Venkatesan, A., \& Shull, J.~M.\ 2013, \apj, 770, 1, 76

\bibitem[Bertoldi \& McKee(1992)]{1992ApJ...395..140B} Bertoldi, F. \& McKee, C.~F.\ 1992, \apj, 395, 140
  
\bibitem[Bigiel et al.(2008)]{2008AJ....136.2846B} Bigiel, F., Leroy, A., Walter, F., et al.\ 2008, \aj, 136, 6, 2846

\bibitem[Bigiel et al.(2011)]{2011ApJ...730L..13B} Bigiel, F., Leroy, A.~K., Walter, F., et al.\ 2011, \apjl, 730, 2, L13
  
\bibitem[Bolatto et al.(2013)]{2013ARA&A..51..207B} Bolatto, A.~D., Wolfire, M., \& Leroy, A.~K.\ 2013, ARA\&A, 51, 207

\bibitem[Condon(1992)]{1992ARA&A..30..575C} Condon, J.~J.\ 1992, \araa, 30, 575

\bibitem[Condon \& Ransom(2016)]{2016era..book.....C} Condon, J.~J. \& Ransom, S.~M.\ 2016, Essential Radio Astronomy
  
\bibitem[Cormier et al.(2014)]{2014A&A...564A.121C} Cormier, D., Madden, S.~C., Lebouteiller, V., et al.\ 2014, \aap, 564, A121

\bibitem[Cormier et al.(2015)]{2015A&A...578A..53C} Cormier, D., Madden, S.~C., Lebouteiller, V., et al.\ 2015, \aap, 578, A53
  
\bibitem[Dale et al.(2012)]{2012ApJ...745...95D} Dale, D.~A., Aniano, G., Engelbracht, C.~W., et al.\ 2012, ApJ, 745, 95 
  
\bibitem[de los Reyes \& Kennicutt(2019)]{2019ApJ...872...16D} de los Reyes, M.~A.~C. \& Kennicutt, R.~C.\ 2019, \apj, 872, 1, 16

\bibitem[Dib(2011)]{2011ApJ...737L..20D} Dib, S.\ 2011, \apjl, 737, 1, L20

\bibitem[Downes \& Solomon(1998)]{1998ApJ...507..615D} Downes, D. \& Solomon, P.~M.\ 1998, \apj, 507, 2, 615
  
\bibitem[Draine et al.(2007)]{2007ApJ...663..866D} Draine, B.~T., Dale, D.~A., Bendo, G., et al.\ 2007, \apj, 663, 2, 866

\bibitem[\protect\citeauthoryear{Elmegreen}{1989}]{Elmegreen89} Elmegreen, B.G. 1989, 338, 178
  
\bibitem[Fernandez \& Shull(2011)]{2011ApJ...731...20F} Fernandez, E.~R. \& Shull, J.~M.\ 2011, \apj, 731, 1, 20
  
\bibitem[Filho et al.(2019)]{2019MNRAS.484..543F} Filho, M.~E., Tabatabaei, F.~S., S{\'a}nchez Almeida, J., et al.\ 2019, \mnras, 484, 1, 543

\bibitem[Fisher et al.(2019)]{2019ApJ...870...46F} Fisher, D.~B., Bolatto, A.~D., White, H., et al.\ 2019, \apj, 870, 1, 46

\bibitem[Freundlich et al.(2019)]{2019A&A...622A.105F} Freundlich, J., Combes, F., Tacconi, L.~J., et al.\ 2019, \aap, 622, A105
  
\bibitem[Gatto et al.(2013)]{2013MNRAS.433.2749G} Gatto, A., Fraternali, F., Read, J.~I., et al.\ 2013, \mnras, 433, 4, 2749

\bibitem[Gardan et al.(2007)]{2007A&A...473...91G} Gardan, E., Braine, J., Schuster, K.~F., et al.\ 2007, \aap, 473, 1, 91
  
\bibitem[Gnedin \& Draine(2014)]{2014ApJ...795...37G} Gnedin, N.~Y. \& Draine, B.~T.\ 2014, \apj, 795, 1, 37

\bibitem[Gratier et al.(2010)]{2010A&A...512A..68G} Gratier, P., Braine, J., Rodriguez-Fernandez, N.~J., et al.\ 2010, \aap, 512, A68
  
\bibitem[Gratier et al.(2017)]{2017A&A...600A..27G} Gratier, P., Braine, J., Schuster, K., et al.\ 2017, \aap, 600, A27

\bibitem[G{\"u}rkan et al.(2018)]{2018MNRAS.475.3010G} G{\"u}rkan, G., Hardcastle, M.~J., Smith, D.~J.~B., et al.\ 2018, \mnras, 475, 3, 3010
  
\bibitem[Hersant et al.(2009)]{2009A&A...493L..49H} Hersant, F., Wakelam, V., Dutrey, A., Guilloteau, S., \& Herbst, E.\ 2009, A\&A, 493, L49 

\bibitem[Huang \& Kauffmann(2014)]{2014MNRAS.443.1329H} Huang, M.-L. \& Kauffmann, G.\ 2014, \mnras, 443, 2, 1329
  
\bibitem[Hunt et al.(2015)]{2015A&A...583A.114H} Hunt, L.~K., Garc{\'\i}a-Burillo, S., Casasola, V., et al.\ 2015, \aap, 583, A114

\bibitem[Kelly(2007)]{2007ApJ...665.1489K} Kelly, B.~C.\ 2007, \apj, 665, 1489
  
\bibitem[Kennicutt et al.(2011)]{2011PASP..123.1347K} Kennicutt, R.~C., Calzetti, D., Aniano, G., et al.\ 2011, \pasp, 123, 910, 1347
  
\bibitem[Kennicutt \& De Los Reyes(2021)]{2021ApJ...908...61K} Kennicutt, R.~C. \& De Los Reyes, M.~A.~C.\ 2021, \apj, 908, 1, 61

\bibitem[Klein et al.(1984)]{1984A&A...141..241K} Klein, U., Wielebinski, R., \& Thuan, T.~X.\ 1984, \aap, 141, 241
  
\bibitem[\protect\citeauthoryear{Kregel et al.}{2002}]{Kregel} Kregel, M., van der Kruit, P.~C., de Grijs, R. 2002, MNRAS, 334, 646
  
\bibitem[Krumholz et al.(2008)]{2008ApJ...689..865K} Krumholz, M.~R., McKee, C.~F., \& Tumlinson, J.\ 2008, \apj, 689, 2, 865

\bibitem[Krumholz et al.(2009)]{2009ApJ...693..216K} Krumholz, M.~R., McKee, C.~F., \& Tumlinson, J.\ 2009, \apj, 693, 1, 216
  
\bibitem[Lacki et al.(2010)]{2010ApJ...717....1L} Lacki, B.~C., Thompson, T.~A., \& Quataert, E.\ 2010, ApJ, 717, 1

\bibitem[Larson(1981)]{1981MNRAS.194..809L} Larson, R.~B.\ 1981, MNRAS, 194, 809
  
\bibitem[Leitherer et al.(2016)]{2016ApJ...823...64L} Leitherer, C., Hernandez, S., Lee, J.~C., et al.\ 2016, \apj, 823, 1, 64

\bibitem[Leroy et al.(2008)]{2008AJ....136.2782L} Leroy, A.~K., Walter, F., Brinks, E., et al.\ 2008, \aj, 136, 6, 2782
  
\bibitem[Leroy et al.(2009)]{2009AJ....137.4670L} Leroy, A.~K., Walter, F., Bigiel, F., et al.\ 2009a, \aj, 137, 6, 4670

\bibitem[Leroy et al.(2009)]{2009ApJ...702..352L} Leroy, A.~K., Bolatto, A., Bot, C., et al.\ 2009b, \apj, 702, 1, 352

\bibitem[Leroy et al.(2011)]{2011ApJ...737...12L} Leroy, A.~K., Bolatto, A., Gordon, K., et al.\ 2011, \apj, 737, 1, 12

\bibitem[Leroy et al.(2022)]{2022ApJ...927..149L} Leroy, A.~K., Rosolowsky, E., Usero, A., et al.\ 2022, \apj, 927, 2, 149

\bibitem[Liszt(2007)]{2007A&A...476..291L} Liszt, H.~S.\ 2007, \aap, 476, 1, 291
  
\bibitem[Liz{\'e}e et al.(2022)]{2022A&A...663A.152L} Liz{\'e}e, T., Vollmer, B., Braine, J., et al.\ 2022, \aap, 663, A152

\bibitem[Luo et al.(2023)]{2023ApJ...946...91L} Luo, G., Zhang, Z.-Y., Bisbas, T.~G., et al.\ 2023, \apj, 946, 2, 91
  
\bibitem[Madden et al.(2013)]{2013PASP..125..600M} Madden, S.~C., R{\'e}my-Ruyer, A., Galametz, M., et al.\ 2013, \pasp, 125, 928, 600
  
\bibitem[Madden \& Cormier(2019)]{2019IAUS..344..240M} Madden, S.~C. \& Cormier, D.\ 2019, Dwarf Galaxies: From the Deep Universe to the Present, 344, 240
  
\bibitem[Madden et al.(2020)]{2020A&A...643A.141M} Madden, S.~C., Cormier, D., Hony, S., et al.\ 2020, \aap, 643, A141

\bibitem[McQuade et al.(1995)]{1995ApJS...97..331M} McQuade, K., Calzetti, D., \& Kinney, A.~L.\ 1995, \apjs, 97, 331
  
\bibitem[Moln{\'a}r et al.(2021)]{2021MNRAS.504..118M} Moln{\'a}r, D.~C., Sargent, M.~T., Leslie, S., et al.\ 2021, MNRAS, 504, 118

\bibitem[Ostriker \& Kim(2022)]{2022ApJ...936..137O} Ostriker, E.~C. \& Kim, C.-G.\ 2022, \apj, 936, 137
  
\bibitem[Pacholczyk(1970)]{1970ranp.book.....P} Pacholczyk, A.~G.\ 1970, Radio astrophysics. Nonthermal processes in galactic and extragalactic sources, Series of Books in Astronomy and Astrophysics

\bibitem[Padoan et al.(1997)]{1997MNRAS.288..145P} Padoan, P., Nordlund, A., \& Jones, B.~J.~T.\ 1997, \mnras, 288, 1
  
\bibitem[Pohl(1993)]{1993A&A...270...91P} Pohl, M.\ 1993, \aap, 270, 91
  
\bibitem[Ramambason et al.(2024)]{2024A&A...681A..14R} Ramambason, L., Lebouteiller, V., Madden, S.~C., et al.\ 2024, \aap, 681, A14

\bibitem[R{\'e}my-Ruyer et al.(2014)]{2014A&A...563A..31R} R{\'e}my-Ruyer, A., Madden, S.~C., Galliano, F., et al.\ 2014, \aap, 563, A31

\bibitem[R{\'e}my-Ruyer et al.(2015)]{2015A&A...582A.121R} R{\'e}my-Ruyer, A., Madden, S.~C., Galliano, F., et al.\ 2015, \aap, 582, A121

\bibitem[Roychowdhury \& Chengalur(2012)]{2012MNRAS.423L.127R} Roychowdhury, S. \& Chengalur, J.~N.\ 2012, \mnras, 423, 1, L127
  
\bibitem[Roychowdhury et al.(2015)]{2015MNRAS.449.3700R} Roychowdhury, S., Huang, M.-L., Kauffmann, G., et al.\ 2015, \mnras, 449, 4, 3700

\bibitem[Roychowdhury et al.(2017)]{2017A&A...608A..24R} Roychowdhury, S., Chengalur, J.~N., \& Shi, Y.\ 2017, \aap, 608, A24

\bibitem[Ruaud et al.(2016)]{2016MNRAS.459.3756R} Ruaud, M., Wakelam, V., \& Hersant, F.\ 2016, \mnras, 459, 4, 3756. doi:10.1093/mnras/stw887

\bibitem[Rujopakarn et al.(2013)]{2013ApJ...767...73R} Rujopakarn, W., Rieke, G.~H., Weiner, B.~J., et al.\ 2013, \apj, 767, 1, 73

\bibitem[Saintonge et al.(2011)]{2011MNRAS.415...32S} Saintonge, A., Kauffmann, G., Kramer, C., et al.\ 2011a, \mnras, 415, 1, 32

\bibitem[Saintonge et al.(2011)]{2011MNRAS.415...61S} Saintonge, A., Kauffmann, G., Wang, J., et al.\ 2011b, \mnras, 415, 1, 61
  
\bibitem[Schinnerer \& Leroy(2024)]{2024ARA&A..62..369S} Schinnerer, E. \& Leroy, A.~K.\ 2024, \araa, 62, 1, 369

\bibitem[Schruba et al.(2012)]{2012AJ....143..138S} Schruba, A., Leroy, A.~K., Walter, F., et al.\ 2012, \aj, 143, 6, 138

\bibitem[Smith et al.(2021)]{2021A&A...648A...6S} Smith, D.~J.~B., Haskell, P., G{\"u}rkan, G., et al.\ 2021, \aap, 648, A6

\bibitem[Tacconi et al.(2013)]{2013ApJ...768...74T} Tacconi, L.~J., Neri, R., Genzel, R., et al.\ 2013, \apj, 768, 1, 74
  
\bibitem[Teich et al.(2016)]{2016ApJ...832...85T} Teich, Y.~G., McNichols, A.~T., Nims, E., et al.\ 2016, \apj, 832, 1, 85

\bibitem[van der Tak et al.(2007)]{2007A&A...468..627V} van der Tak, F.~F.~S., Black, J.~H., Sch{\"o}ier, F.~L., Jansen, D.~J., \& van Dishoeck, E.~F.\ 2007, A\&A, 468, 627

\bibitem[Vieu et al.(2022)]{2022MNRAS.512.1275V} Vieu, T., Gabici, S., Tatischeff, V., et al.\ 2022, \mnras, 512, 1275
 
\bibitem[Vollmer \& Leroy(2011)]{2011AJ....141...24V} Vollmer, B., \& Leroy, A.~K.\ 2011, AJ, 141, 24 

\bibitem[Vollmer et al.(2017)]{2017A&A...602A..51V} Vollmer, B., Gratier, P., Braine, J., et al.\ 2017, \aap, 602, A51

\bibitem[Vollmer et al.(2022)]{2022A&A...667A..30V} Vollmer, B., Soida, M., \& Dallant, J.\ 2022, \aap, 667, A30

\bibitem[Vollmer et al.(2025)]{2025A&A...693A.267V} Vollmer, B., Freundlich, J., Gratier, P., et al.\ 2025, \aap, 693, A267

\bibitem[Wakelam et al.(2012)]{2012ApJS..199...21W} Wakelam, V., Herbst, E., Loison, J.-C., et al.\ 2012, \apjs, 199, 21

\bibitem[Wang et al.(2019)]{2019A&A...631A.109W} Wang, L., Gao, F., Duncan, K.~J., et al.\ 2019, \aap, 631, A109
  
\bibitem[Wolfire et al.(2010)]{2010ApJ...716.1191W} Wolfire, M.~G., Hollenbach, D., \& McKee, C.~F.\ 2010, \apj, 716, 2, 1191
  
\bibitem[Wyder et al.(2009)]{2009ApJ...696.1834W} Wyder, T.~K., Martin, D.~C., Barlow, T.~A., et al.\ 2009, \apj, 696, 2, 1834
  
\bibitem[Yun et al.(2001)]{2001ApJ...554..803Y} Yun, M.~S., Reddy, N.~A., \& Condon, J.~J.\ 2001, ApJ, 554, 803 
  
\end{thebibliography}
\end{document}